\documentclass[twocolumn]{aastex62}

\usepackage{graphicx}	
\usepackage{amsmath}	
\usepackage{amssymb}	
\usepackage{xspace}

\usepackage{siunitx}
\usepackage{comment}

\newcommand{\gDor}{$\gamma$~Dor\xspace}
\newcommand{\gmodes}{g~modes\xspace}
\newcommand{\Msun}{\,M$_{\odot}$\xspace}
\newcommand{\dMsun}{-M$_{\odot}$\xspace}
\newcommand{\kms}{~{\rm km~s}$^{-1}$\xspace}

\newcommand{\el}[2]{$^{#1}{\rm #2}$\xspace}
\newcommand{\grad}{$g_{{\rm rad}, i}$\xspace}
\newcommand{\xc}{$X_{\rm c}$\xspace}
\newcommand{\zini}{$Z_{\rm ini}$\xspace}
\newcommand{\fov}{$f_{\rm ov}$\xspace}
\newcommand{\frot}{$f_{\rm rot}$\xspace}
\newcommand{\Prot}{$P_{\rm rot}$\xspace}

\newcommand{\teff}{$T_{\rm eff}$\xspace}
\newcommand{\logg}{$\log g$\xspace}
\newcommand{\vsini}{$v \sin i$\xspace}

\newcommand{\pin}{$\Pi_0$\xspace}
\newcommand{\ko}{KIC\,11145123\xspace}
\newcommand{\kn}{KIC\,9751996\xspace}
\newcommand{\comm}[1]{#1}
\newcommand{\car}[1]{#1}
\newcommand{\cart}[1]{#1}
\newcommand{\Table}[1]{Table~#1\xspace}
\newcommand{\mesa}{\texttt{MESA}\xspace}
\newcommand{\gyre}{\texttt{GYRE}\xspace}
\newcommand{\chir}{$\chi^2_{\rm red}$\xspace}

\graphicspath{{./}{figures/}}

\received{2020 Jan 13}
\revised{2020 Mar 27}
\accepted{2020 Apr 23}
\submitjournal{ApJ}

\shorttitle{Atomic diffusion in \gDor pulsators}
\shortauthors{Mombarg et al.}

\begin{document}

\title{Asteroseismic modeling of gravity modes in slowly rotating A/F stars with radiative levitation}

\correspondingauthor{Joey Mombarg}
\email{joey.mombarg@kuleuven.be}

\author[0000-0002-9901-3113]{Joey S. G. Mombarg}
\affil{Institute of Astronomy, KU Leuven, Celestijnenlaan 200D, B-3001 Leuven, Belgium}

\author{Aaron Dotter}
\affiliation{Center for Astrophysics \textbar~Harvard \& Smithsonian, Cambridge, MA 02138, USA}

\author{Timothy Van Reeth}
\affiliation{Institute of Astronomy, KU Leuven, Celestijnenlaan 200D, B-3001 Leuven, Belgium}

\author{Andrew Tkachenko}
\affiliation{Institute of Astronomy, KU Leuven, Celestijnenlaan 200D, B-3001 Leuven, Belgium}

\author{Sarah Gebruers}
\affiliation{Institute of Astronomy, KU Leuven, Celestijnenlaan 200D, B-3001 Leuven, Belgium}

\author{Conny Aerts}
\affiliation{Institute of Astronomy, KU Leuven, Celestijnenlaan 200D, B-3001 Leuven, Belgium}
\affiliation{Department of Astrophysics, IMAPP, Radboud University Nijmegen, PO Box 9010, 6500 GL Nijmegen, The Netherlands}
\affiliation{Max Planck Institute for Astronomy, Koenigstuhl 17, 69117 Heidelberg, Germany}



\begin{abstract}
It has been known for several decades that transport of chemical elements is induced by the process of microscopic atomic diffusion. Yet, the effect of atomic diffusion, including radiative levitation, has hardly been studied in the context of gravity mode pulsations of core-hydrogen burning stars. In this paper, we study the difference in the properties of such modes for models with and without atomic diffusion. We perform asteroseismic modeling of two slowly rotating A- and F-type pulsators, \ko (\frot~$\approx0.010~{\rm d}^{-1}$) and \kn (\frot~$\approx0.0696~{\rm d}^{-1}$), respectively, based on the periods of individual gravity modes. For both stars, we find models whose g-mode periods are in very good agreement with the {\it Kepler\/} asteroseismic data, keeping in mind that the theoretical/numerical precision of present-day stellar evolution models is typically about two orders of magnitude lower than the measurement errors. Using the Akaike Information Criterion (AIC) we have made a comparison between our best models with and without diffusion, and found very strong evidence for signatures of atomic diffusion in the pulsations of \ko. In the case of \kn the models with atomic diffusion are not able to explain the data as well as the models without it. Furthermore, we compare the observed surface abundances with those predicted by the best fitting models. The observed abundances are inconclusive for \kn, while those of \ko from the literature can better be explained by a model with atomic diffusion. 

\end{abstract}



\section{Introduction} \label{sec:intro}
The mechanism(s) driving the transport of angular momentum \citep[e.g.][]{Aerts2019-ARAA} and chemical elements \citep[e.g.][]{Salaris2017} within stars are still not understood from stellar evolution theory. Discrepancies between observations and theory have been shown for stars with birth masses between 1.3 and 8\Msun, which comprise a convective core, enshrouded by a radiative envelope (possibly with internal convective shells from partial ionization zones or a thin outer convective envelope for $M_\star \lesssim 1.6$\Msun). 
In these radiative envelopes, the transport of chemical elements on a macroscopic scale is ascribed to convective core overshooting \citep[e.g.][]{Viallet2015}, rotation \citep[e.g.][]{maeder2009}, or internal gravity waves \citep[IGW, e.g.][]{RogersMcElwaine2017}, whereas \comm{the transport of chemical elements} on a microscopic scale is the result of atomic diffusion \citep{Michaud2015}. In stellar evolution codes, the description of macroscopic mixing introduces additional free parameters, whereas mixing from atomic diffusion can be derived from first principles. So far, the theory of element transport has been mainly evaluated by measurements of surface abundances. Asteroseismology constitutes a novel technique to empirically assess the conditions deep inside the interior of a star \citep{aerts2010}, as well as its evolutionary history \citep[e.g.][]{Bowman2019-nature}. 
The unprecedented high-quality data from the space-based \textit{CoRoT\/} \citep{Auvergne2009}, \textit{Kepler\/} \citep{borucki2010}, and \textit{TESS\/} \citep{Ricker2015} missions allow for scrutiny of the current stellar evolution models of the stars' interiors by means of gravito-inertial asteroseismology \citep{Aerts2018-apjs}. The current state-of-the-art stellar models and pulsation codes are not capable of reproducing the observed oscillation frequencies of gravity (g) modes in $\gamma$ Doradus (\gDor, cf. \cite{Kurtz2014, Saio2015, VanReeth2016, Schmid2016}) and Slowly-Pulsating B-type (SPB, cf. \cite{papics2014, Moravveji2015, szewczuk2018, Aerts2019-CoRoT}) stars within the uncertainties of the data. Hence, additional physics is required in order to improve both the current stellar models as well as the prediction of the g-mode frequencies from these equilibrium models. Studies have already demonstrated the potential of \gmodes to distinguish between different near-core mixing profiles \citep{Pedersen2018} and the temperature gradient close to the convective core interface \citep{Michielsen2019}.  The work of \cite{Aerts2018-apjs} has evaluated a hierarchy of input physics when modeling \gmodes across a wide mass range. In the current work, we investigate to what extent the process of atomic diffusion can improve the theoretically predicted oscillation frequencies in two slowly rotating \gDor stars. 

Throughout this paper, we use the umbrella term `atomic diffusion' to refer to the following four diffusion processes; \textit{(i)} gravitational settling, \textit{(ii)} thermal diffusion, \textit{(iii)} concentration diffusion, and \textit{(iv)} radiative levitation. The former process causes elements heavier than hydrogen to migrate towards the stellar center. This process of gravitational settling is counteracted by the process of radiative levitation where momentum is transferred from the radiative flux -- generated in the core -- to the atoms, thereby `levitating' them outwards. The cross-section for photon absorption is larger for heavier elements and therefore radiative levitation will be most dominant on the heaviest elements, while the lighter elements will primarily be subjected to gravitational settling. Thermal diffusion works in the same direction as gravitational settling as heavier elements are pulled towards hotter regions due to the interaction with field protons \citep{Michaud2015}. The importance of atomic diffusion has already been pointed out in context of helioseismology \citep{christensen-dalsgaard1993}, chemical tagging \citep{Dotter2017}, and helium abundance determination \citep[][albeit without radiative levitation]{Verma2019}. Yet, when it comes to asteroseismic modeling of g-mode pulsators, this process is not taken into account as it presents a computationally challenging task. Since \gmodes probe the stellar structure in the radiative region of stars where diffusive processes are at work, their frequencies will be dependent on the treatment of these processes in this region. The effect of atomic diffusion, especially radiative levitation, imposes implications on the determination of the properties of solar-like oscillators, as has been demonstrated by \cite{Deal2018, Deal2019}. These authors focused their investigations on pressure (p) modes in solar-like oscillators, with masses up to 1.44\Msun. The dominant restoring force for p~mode oscillations is the pressure force, whereas in this work we are concerned with gravito-inertial modes, for which the buoyancy and Coriolis force both act as restoring forces. For \gDor stars, the mass regime is about 1.4\Msun to 1.9\Msun and thus covers a slightly higher mass regime where the force of the radiative levitation will be more dominant. Furthermore, the frequencies of \gmodes are dependent on the behavior of the chemical gradient in the near-core region. We aim to characterize the impact of atomic diffusion on the chemical gradient, and the implications for the g-mode frequencies.

Atomic diffusion might also be a key ingredient in the excitation of modes as it introduces accumulation of iron and nickel in the stellar layer where the opacity reaches its maximum \citep{Richard2001, Deal2016, HuiBonHoa2018}, thereby altering the Rosseland mean opacity. This so-called `opacity bump' occurs around a temperature of \SI{200000}{\K}, i.e., the ionization temperature of iron. This may result in the forming of extra convection zones, which can excite pulsations through the $\kappa$-mechanism. \cite{Michaud2015} show these iron convection zones appear and disappear throughout the main-sequence lifetime for a 1.5\dMsun model, but are persistent for a 1.7\dMsun and a 1.9\dMsun model. 

In this work we will focus our attention on the theoretically predicted adiabatic frequency values and compare them with those of observed modes. We take a data-driven approach and test our improved stellar models against observations from two stars observed by the nominal \textit{Kepler\/} mission.
As the effects of atomic diffusion might be (partially) washed out by rotationally induced mixing, we have selected two slowly rotating \gDor stars for our purposes; \ko and \kn. The former has been studied by \cite{Kurtz2014}, who found 15 g-mode triplets ($m=0$ modes with low visibility), one p-mode triplet, and one p-mode quintuplet. From the frequency splittings, a nearly uniform rotation period of $\sim$\SI{100}{\day} (upper limit of \frot~=~$0.009512 \pm 0.000002~{\rm d^{-1}}$ from \gmodes) was inferred. Such a slow rotation period is quite rare compared to observations from the study by \cite{Li2019}, who show that the rotation-period distribution -- based on 611 \gDor stars -- is Gaussian-like and peaks around $\sim$\SI{1}{\day}, although some excess around slow rotation is seen in the distribution. A spectroscopic follow-up of \ko by \cite{Takada-Hidai2017} shows the star has a low surface metallicity ([Fe/H] = $-0.71\,\pm\,0.11$~dex) and its low lithium abundance is compatible with those of the blue stragglers \citep{Glaspey1994}. Moreover, these authors did not find any presence of a companion. In this paper, we test whether the observed pulsation periods and surface abundances can be explained by a single star model with atomic diffusion taken into account.
The second star, \kn, has a super-solar surface metallicity, [Fe/H] = +0.28 $\pm$ 0.07 dex \citep{VanReeth2015}. Also for this star rotationally split modes were identified, yielding a rotation frequency of 0.0696 $\pm$ \SI{ 0.0008 }{\per \day} (\Prot = 14.4 $\pm$ \SI{ 1.7}{\day}, \cite{VanReeth2016}). \kn is one of the 37 \gDor stars in the sample analyzed by \cite{Mombarg2019}. In this work, we quantify the difference in inferred mass and age from asteroseismic modeling of the observed g-mode frequencies of both \ko and \kn, using models with and without atomic diffusion.

\section{Methods} \label{sec:methods}
The \comm{accelerations due to radiative levitation} in the stellar interior can be computed either by means of `opacity sampling' \citep[e.g.][]{LeBlanc2000} or by using the Single-Valued Parameter (SVP) method \citep[cf.][]{Alecian2002, LeBlanc2004, Theado2012, Deal2019}. In this work, we have used the opacity sampling method, which is a slightly more accurate method compared to the SVP method, but this increased accuracy comes at the cost of computation time.
The inclusion of radiative levitation drastically increases the computation time of an evolution track, e.g., an increase by about a factor 100 (using four threads) when starting at the pre-main sequence (pre-MS) contraction up to the terminal age main-sequence (TAMS). Hence, we employ the method from \cite{Mombarg2019} as a first estimate for the mass range of the grids to be constructed to model the individual \comm{periods} from the effective temperature (\teff), surface gravity (\logg), and the reduced asymptotic period spacing (\pin). For the determination of \pin for \kn, we rely on the work by \cite{VanReeth2016}. Since the rotation period of \ko is extremely long, we estimate \pin by taking the average period spacing of the central frequencies found by \cite{Kurtz2014} (0.0241 $\pm$ 0.0009~d), and use \pin$ \approx \Delta P_{\rm co} \sqrt{l(l+1)}$. The period spacing in the corotating frame $\Delta P_{\rm co}$ is in this case roughly equal to the period spacing in the inertial frame. The coverages in metallicity for the two stars are based on spectroscopic measurements from \cite{Takada-Hidai2017} and \cite{VanReeth2016} for \ko and \kn, respectively, where the measured value and the upper and lower values of the uncertainty intervals listed in these studies are used. Besides the mass, the method by \cite{Mombarg2019} also provides us with an estimate for the hydrogen mass fraction inside the convective core (\xc). These estimates indicate both stars have \xc$\lesssim 0.3$. Following Fig. 8 of \cite{Deal2018}, the measured [Fe/H] should not differ more than 0.1~dex from the initial value. Therefore, the observed iron surface abundance will most likely be a good approximation of the star's initial bulk metallicity, \zini. Nevertheless, we have also tested if the pulsations can be explained by models with atomic diffusion at solar metallicity. \Table{\ref{tab:grids}} summarizes the extent of the grids used in this work (the grid where \zini is fixed at 0.014 covers the same ranges for the other input parameters). 

\begin{table}[]
    \centering
    \begin{tabular}{cccc}
    Parameter & Lower& Upper & Step size \\
     & boundary & boundary &  \\

    \hline
    \multicolumn{4}{c}{\bf \kn}\\
       $M_\star$   & 1.65\Msun & 1.90\Msun & 0.01\Msun \\
       $Z_{\rm ini}$  & 0.022 & 0.030 & 0.004 \\
       $f_{\rm ov}$ & \multicolumn{3}{c}{$\in$ [0.0100, 0.0175, 0.0225, 0.0300]} \\
        $f_{\rm rot}$ & \multicolumn{3}{c}{fixed at \SI{0.0696}{\per \day}} \\
           \multicolumn{4}{c}{\bf \ko} \\
       $M_\star$   & 1.30\Msun & 1.50\Msun & 0.01\Msun \\
       $Z_{\rm ini}$  & 0.002 & 0.004 & 0.001 \\
       $f_{\rm ov}$ & \multicolumn{3}{c}{$\in$ [0.0100, 0.0175, 0.0225, 0.0300]} \\ 
       $f_{\rm rot}$ & \multicolumn{3}{c}{fixed at \SI{0.010}{\per \day}} \\

    \end{tabular}
    \caption{Extend of the grids used to model \kn and \ko. Grids are computed for {\it (i)} standard OP opacities, without atomic diffusion, and {\it(ii)} OP monochromatic opacities with atomic diffusion.}
    \label{tab:grids}
\end{table}
\subsection{Stellar models}
The stellar models were computed using the stellar evolution code \mesa \citep{Paxton2011, Paxton2013, Paxton2015, Paxton2018, Paxton2019}, r11701. For both the grid with and without atomic diffusion the \comm{opacities} from the Opacity Project (OP; \cite{Seaton2005}) were used, where we relied on the monochromatic opacities for the computations of the \comm{accelerations due to radiative levitation}. \comm{The Rosseland mean opacity is computed from the monochromatic opacities for models with atomic diffusion.}
We point out that the computation time of models using the OP monochromatic opacity tables has been improved compared to \mesa revisions released prior to \mesa Paper V \citep{Paxton2019}, and these models also benefit from multi-treading. 
A solar mixture according to \cite{Asplund2009} is assumed, as spectroscopic studies of \gDor stars have shown that these stars have abundance patterns similar to that of the Sun \citep[e.g.][]{Kahraman2016}. From this mixture the metal abundances are scaled with \zini. 

The boundary of the convective core is defined according to the Ledoux criterion and we impose diffusive convective core overshooting. The core boundary mixing (CBM) efficiency, $D_{\rm CBM}$, is then parameterized as per \cite{Freytag1996} by
\begin{equation}
    D_{\rm CBM}(r) = D_{\rm CBM}(r_0)\exp\left(\frac{-2(r - r_0)}{f_{\rm ov}H_{\rm P}(r_{\rm cb})}\right),
    \label{eq:D_CBM}
\end{equation}
where $D_{\rm CBM}(r_0)$ is the mixing coefficient at the starting point of the overshoot profile, and $H_{\rm P}(r_{\rm cb})$ the pressure scale height at the convective boundary radius $r_{\rm cb}$. The starting point of the overshoot profile $r_0$ is placed at a radius $r_{\rm cb} - f_0H_{\rm P}(r_{\rm cb})$, since $D_{\rm CBM}(r)$ drops off steeply at the core boundary, making it non-trivial to define its value precisely at $r_{\rm cb}$. In this work, we adopted a value $f_0 = 0.005$. \car{The mixing at the boundary of convective shells and at the bottom of the surface convection zone is described by the same formalism presented in Eq.~(\ref{eq:D_CBM}), for which we set $f_{\rm ov} = 0.015$. }
The basic chemical reaction network in \mesa is extended to include \comm{Ne, Na, Al, Si, S, Ca,} Fe, and Ni. \comm{We considered two ways to set the initial chemical composition $(X_{\rm ini}, Y_{\rm ini}, Z_{\rm ini})$. In a first approach, we fix} the initial hydrogen content $X_{\rm ini} = 0.7154$ and leave the initial metallicity $Z_{\rm ini}$ as a free parameter\comm{, following the larger importance of estimating \zini compared to $Y_{\rm ini}$ for g-mode asteroseismology of intermediate-mass stars \citep{Moravveji2015}}. The initial helium content is then set by $Y_{\rm ini} = 1 - X_{\rm ini} - Z_{\rm ini}$ once \zini is chosen. The \el{2}{H}/\el{1}{H} and \el{3}{He}/\el{4}{He} isotope ratios are set to $2\cdot10^{-5}$ and $1.66\cdot10^{-4}$, respectively, as per \cite{Asplund2009}. \comm{We have chosen this approach to set the initial composition as \cite{Takada-Hidai2017} have estimated asteroseismically that \ko has a high initial helium abundance ($Y_{\rm ini} = 0.297$), compared to the chemical enrichment rate $Y_{\rm ini} = 0.244 + 1.226Z_{\rm ini}$ found by \cite{Verma2019}. In a second approach, we use this enrichment rate by \cite{Verma2019}, despite its large uncertainties, to set the initial composition for both stars. The initial composition of the models presented in this section and in Sec.~\ref{sec:style} is set by fixing $X_{\rm ini}$.}
As atmospheric boundary condition, an Eddington gray atmosphere is used, for which the solar-calibrated mixing length parameter $\alpha_{\rm MLT}$ = 1.713 \citep{Jieun2018}. 
\comm{This value of $\alpha_{\rm MLT}$ has been calibrated with models including atomic diffusion. Omitting atomic diffusion in the stellar evolution models will affect the inferred value of this parameter. The impact of $\alpha_{\rm MLT}$ on the reduced asymptotic period spacing, $\Pi_0$, has been assessed by \cite{johnston2019} and \cite{Mombarg2019}. Both these studies show $\alpha_{\rm MLT}$ to have a small effect on \pin, compared to typical uncertainties for this quantity derived from the {\it Kepler\/} photometry by \cite{VanReeth2016}. }
Furthermore, a small amount of additional mixing is imposed in the radiative zone. A constant mixing efficiency of $D_{\rm mix}(r) = 1.0~{\rm cm^2\,s^{-1}}$ is chosen, following the asteroseismic study by \cite{VanReeth2016}. At the start of each model, a pre-MS model is computed with atomic diffusion already turned on, such that a consistent comparison can be made. No diffusion computations are done in convective zones where the material is assumed to be instantaneously mixed. 

\subsection{Computations of radiative levitations}
\begin{figure}
    \centering
    \includegraphics[width = 0.45\textwidth]{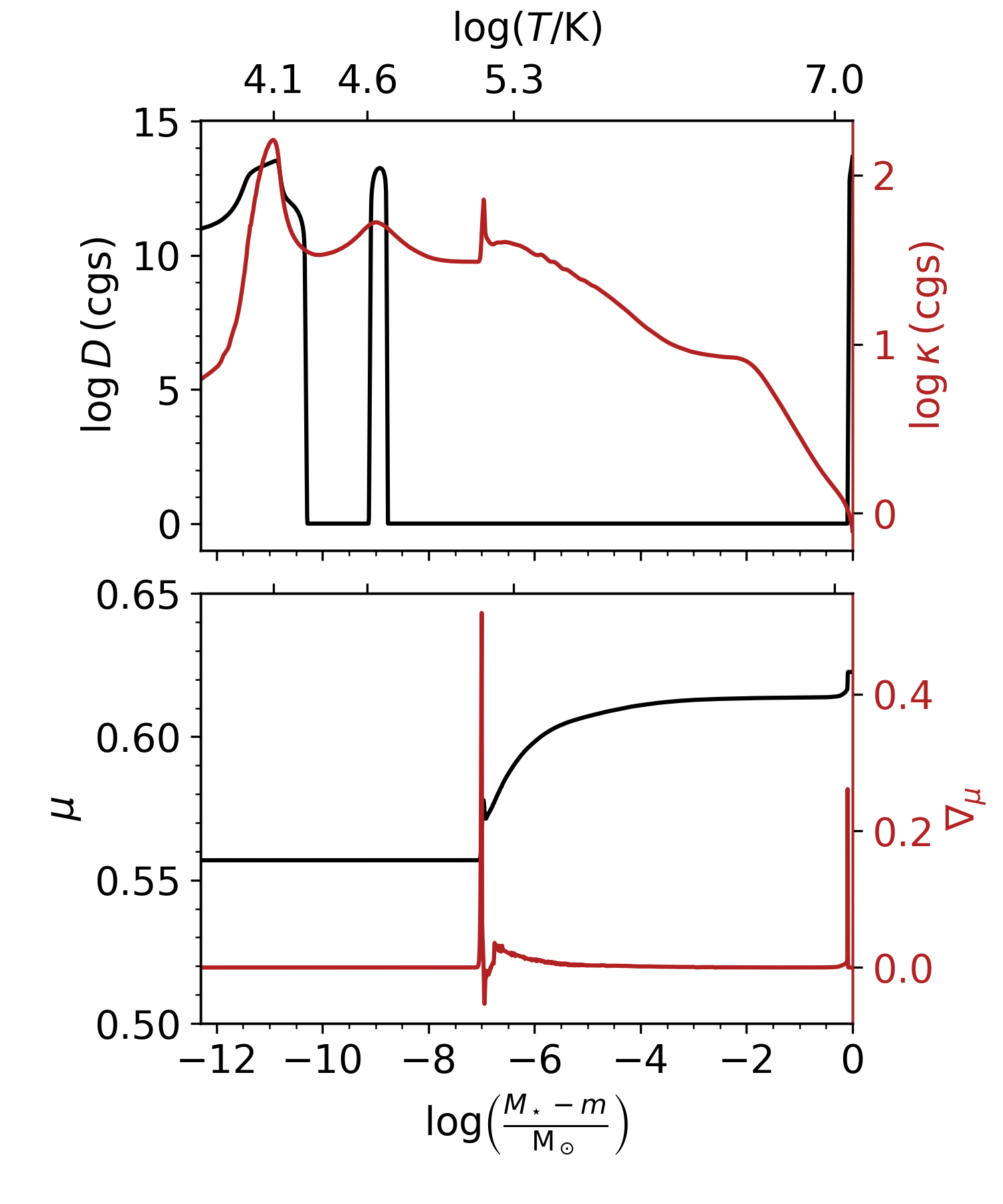}
    \caption{\car{\textit{Top panel:} Profiles of the mixing efficiency (in black) and Rossland mean opacity (in red) for a 1.7\dMsun model ($Z_{\rm ini} = 0.014, f_{\rm ov} = 0.0225$) with an age of 100\,Myr, when atomic diffusion is active. \textit{Bottom panel:} Corresponding profile of the mean molecular weight (fully ionized) $\mu$, and $\nabla_\mu$.}}
    \label{fig:M170_profiles}
\end{figure}

 \comm{Radiative levitations} are computed for the elements \el{1}{H}, \el{3}{He}, \el{4}{He}, \el{12}{C}, \el{14}{N}, \el{16}{O}, \el{20}{Ne}, \el{23}{Na}, \el{24}{Mg}, \el{27}{Al}, \el{28}{Si}, \el{32}{S},  \el{40}{Ca}, \el{56}{Fe}, and \el{58}{Ni}. 
The description of atomic diffusion in \mesa is based on Burgers' equations \citep{Burgers1969}, following the routines by \cite{Thoul1994} for settling and temperature/concentration gradients, 
\begin{equation}
\begin{split}
\frac{p}{K_0}\left(\alpha_i\frac{d \ln p}{dr} + \nu_i\frac{d \ln T}{dr} + \sum_{j=1,\neq e}^{S}\gamma_{ij}\frac{d \ln C_j}{dr}\right) \\
= \sum_{j=1}^{2S+2}\Delta_{ij}W_j,
\label{burgers} 
\end{split}
\end{equation}
that solve for $W_j$, describing the mean diffusion velocity of species $j$,
where $p$ is the pressure, $T$ the temperature, and $C_j$ the concentration of species $j$. The latter is defined as,
\begin{equation}
    C_j = n_j/n_e,
\end{equation}
with $n_j$ and $n_e$ the number densities of the species and electrons, respectively.
The sum is taken over all $S$ species (that includes electrons as well) and $W_j$ is defined as follows,
\[   
W_j = 
     \begin{cases}
       w_j & \text{for}\,j \in [1, S]\\
       r_j & \text{for}\,j \in [S+1, 2S] \\
       K_0^{-1}n_e eE & \text{for}\,j = 2S + 1,\\
       K_0^{-1}n_e m_{\rm p}g & \text{for}\,j = 2S + 2, \\ 
     \end{cases}
\]
where $w_j$ is the mean diffusion velocity and $r_j$ the residual heat flow velocity of a species. We refer to \cite{Thoul1994} for the extensive details, as well as the definition of the constants $K_0$, $\alpha_i$, $\nu_i$, $\gamma_{ij}$ and $\Delta_{ij}$. \mesa uses the diffusion coefficients from \cite{Stanton2016} for the ion-ion terms and the coefficients from \cite{Paquette1986} for electron-ion terms to compute the $\Delta_{ij}$ coefficients. The inclusion of radiative \comm{levitation} in \mesa has been incorporated following \cite{hu2011}, along with some modifications for which we refer to the \mesa Paper III \citep{Paxton2015}. When \comm{accelerations induced by radiative levitation}, \grad, are included, Eq.\,(\ref{burgers}) becomes
\begin{equation}
    \begin{split}
    \frac{p}{K_0} \left(-\frac{\alpha_im_i g_{{\rm rad},i}}{k_{\rm B}T} + \alpha_i\frac{d \ln p}{dr} + \nu_i\frac{d \ln T}{dr} \right. \\ 
    \left. + \sum_{j=1,\neq e}^{S}\gamma_{ij}\frac{d \ln C_j}{dr} \vphantom{\frac{\alpha_im_i g_{{\rm rad},i}}{k_{\rm B}T}} \right) 
    = \sum_{j=1}^{2S+2}\Delta_{ij}W_j,
    \end{split}
\end{equation}
\comm{where $m_{i}$ is the mass of species $i$.
The acceleration induced by radiative levitation} according to \cite{hu2011} is computed as 
\begin{equation}
    g_{{\rm rad},i} = \frac{\mu \kappa_{\rm R}}{\mu_i c} \frac{l(r)}{4 \pi r^2} \int_0^\infty \frac{\sigma_i(u)[1-e^{-u}] - a_i(u)}{\sum_k f_k \sigma_k(u)} {\rm d}u,
\end{equation}
where $\mu$ is the mean molecular weight, $\mu_i$ the molecular weight of species $i$, $c$ is the speed of light, $l(r)$ the luminosity at radius $r$ and $\kappa_{\rm R}$ the Rossland mean opacity. The integral is taken over $u = h\nu/k_{\rm B}T$ ($h$ en $k_{\rm B}$ the Planck and Boltzmann constant, respectively), with $\sigma_i(u)$ the monochromatic opacity cross-section for absorption and scattering, $a_i(u)$ a correction term, and $f_i$ the so-called `number fraction' such that the sum over all species yields the total cross-section of the mixture. In Fig.~\ref{fig:M170_profiles} we show the profiles of the mixing efficiency and the opacity (top panel) as well as the \car{(fully ionized) mean molecular weight $\mu \approx 1/(2X + 3/4Y + Z/2)$ and $\nabla_\mu = {\rm d}\ln\,\mu / {\rm d}\ln\,p$ (bottom panel) for a 1.7\dMsun star. To ensure numerical stability, the outer $10^{-5}$\% of the mass is treated as a single cell for the purposes of atomic diffusion. This explains why the helium convection zone (around $\log\,T = 4.6$) is still present in Fig.~\ref{fig:M170_profiles}, as gravitational settling of helium would otherwise increase the local $\mu$, since $g_{\rm rad, He} \ll g$, thereby stabilizing this convection zone. However, the probing power of g~modes is weak at the stellar surface, compared to the near-core region, and hence, such a numerical approximation has a negligible effect on the asteroseismic modeling performed in this paper. } We do point out \cite{Deal2016} observe an iron convection zone around $\log\,T = 5.3$ for stars $M_\star \gtrsim 1.7{\rm M_\odot}$. Here, we only observe such a convection zone when the opacities are artificially increased with a factor 5 around the Z-bump, in agreement with the study by \cite{guzik2018}. It is not surprising that differences between model properties in the outer envelope resulting from different codes occur, as the computations by \cite{Deal2016} rely on the SVP approximation while \cart{we used the opacity sampling method and treated the $10^{-5}$\% of outer mass as a single cell for atomic diffusion.} Such differences in the properties of the outer envelope are not important for g-mode asteroseismology, because the kernels of such modes probe the deep interior layers of the stars and not the outer envelope.   

\subsection{Pulsation computations}
For each stellar model in the grids, the theoretical pulsation frequencies are computed in the adiabatic approximation using the stellar pulsation code \gyre \citep{Townsend2013, Townsend2018}, version 5.2. For both \ko and \kn, dipole modes have been observed. For both stars, we fix the rotation frequency to the measured value (cf. Table~\ref{tab:grids}), as the near-core rotation rates have been precisely determined in a model-independent way from mode splitting. As both stars are slow rotators -- rotation frequency much smaller than 10$\%$ of the Roche critical rotation -- it is justified to use the traditional approximation of rotation (TAR, \cite{eckart1960}) to compute the frequencies of the \gmodes. Here, the effect of only the (local) vertical component of the Coriolis force is taken into account \citep[cf.][]{Ouazzani2018}. \cite{Li2018} have shown that the typical range of excited radial orders for \gDor stars spans from about -20 to -100. Yet, we limit the computations to a radial order range between $n = -15$ and $n = -45$, based on the identified orders by \cite{Kurtz2014} and \cite{VanReeth2016}, for \ko and \kn, respectively.

\section{The effect of atomic diffusion} \label{sec:style}
\begin{figure}
    \centering
    \includegraphics[width = 0.5\textwidth]{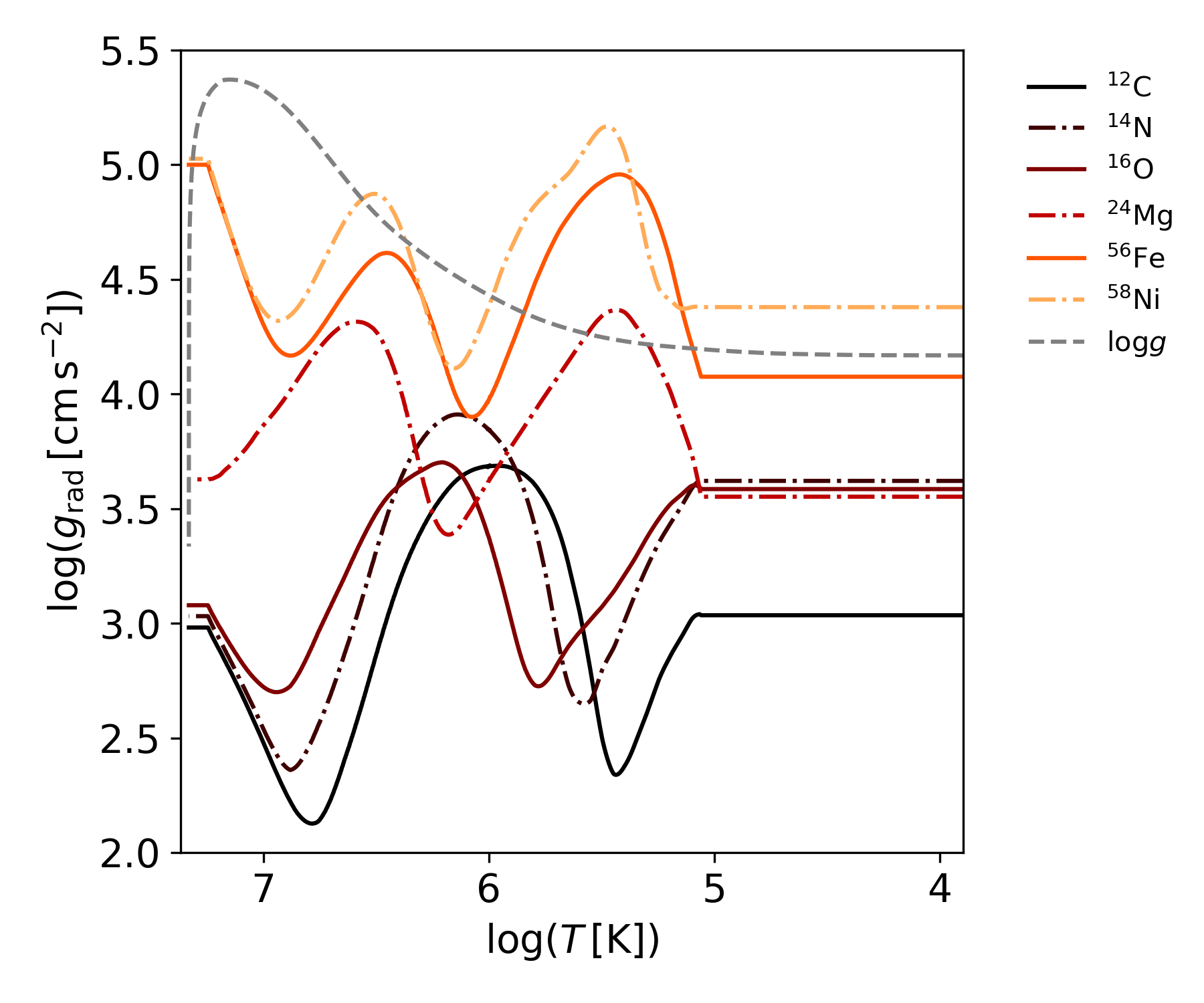}
    \caption{\comm{Accelerations induced by radiative levitation} of several heavy elements in a 1.7\dMsun star at \xc = 0.50 (\zini = 0.014, \fov = 0.0175). The gravitational acceleration is indicated by the dashed gray line. For numerical purposes \grad is kept constant in the outermost cells.  }
    \label{fig:grad}
\end{figure}

Within the mass regime covered by \gDor stars, the forces on the chemical elements due to gravity and radiative levitation are competing with each other \citep{Turcotte1998,Deal2019}. In Fig. \ref{fig:grad}, we show the \grad for a few heavy elements, compared to the gravitational acceleration in a 1.7\dMsun star.  We see that it is important that both these diffusion processes are taken into account for this mass regime, as only including gravitational settling yields unphysical surface abundances \citep{Morel2002}. However, the counteracting role of radiative levitation may also be (partly) taken up by additional mixing processes such as turbulent diffusion \citep[e.g.][]{Dotter2017}, rotational mixing \citep[e.g.][]{Deal2019} or IGW. For the elements plotted in Fig.~\ref{fig:grad}, we report that the order of magnitude and global behavior of \grad in our \mesa model is in agreement with the 1.7\dMsun model computed with the Montreal/Montpellier code by \cite{Richard2001}. Contrary to \mesa, the Montreal/Montpellier code uses the monochromatic opacities from the OPAL project. \newline

\begin{figure}
    \centering
    \includegraphics[width = 0.45\textwidth]{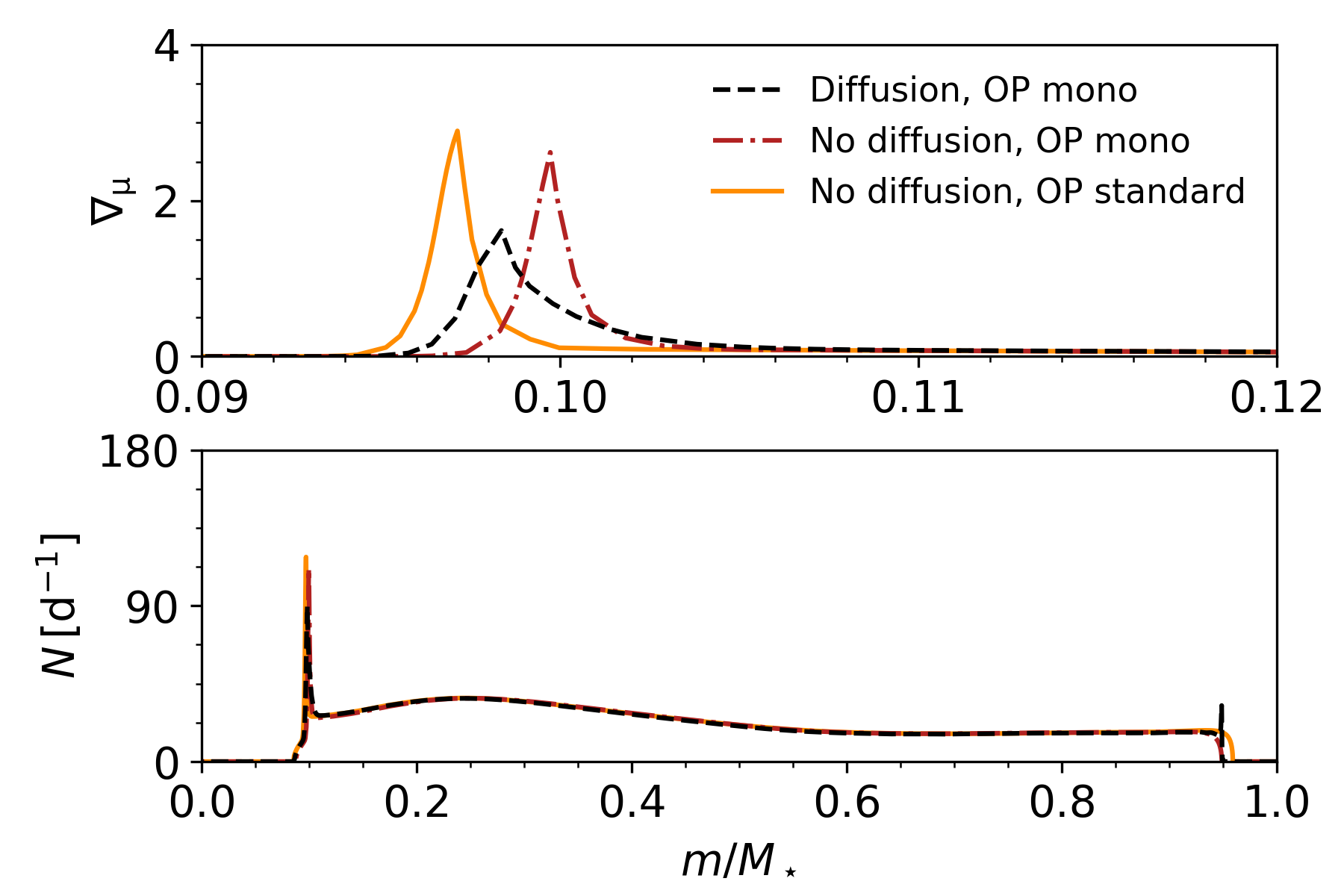}
    \caption{The effect of atomic diffusion on the chemical gradient (top panel) and the Brunt-V\"ais\"al\"a frequency (bottom panel) in a 1.4\dMsun star around \xc = 0.6 with solar metallicity (\zini = 0.014). A comparison is shown for a model computed with the standard OP tables and one using the monochromatic OP opacities. }
    \label{fig:effect_N+mu}
\end{figure}

 \begin{figure*}
    \centering
    \includegraphics[width = \linewidth]{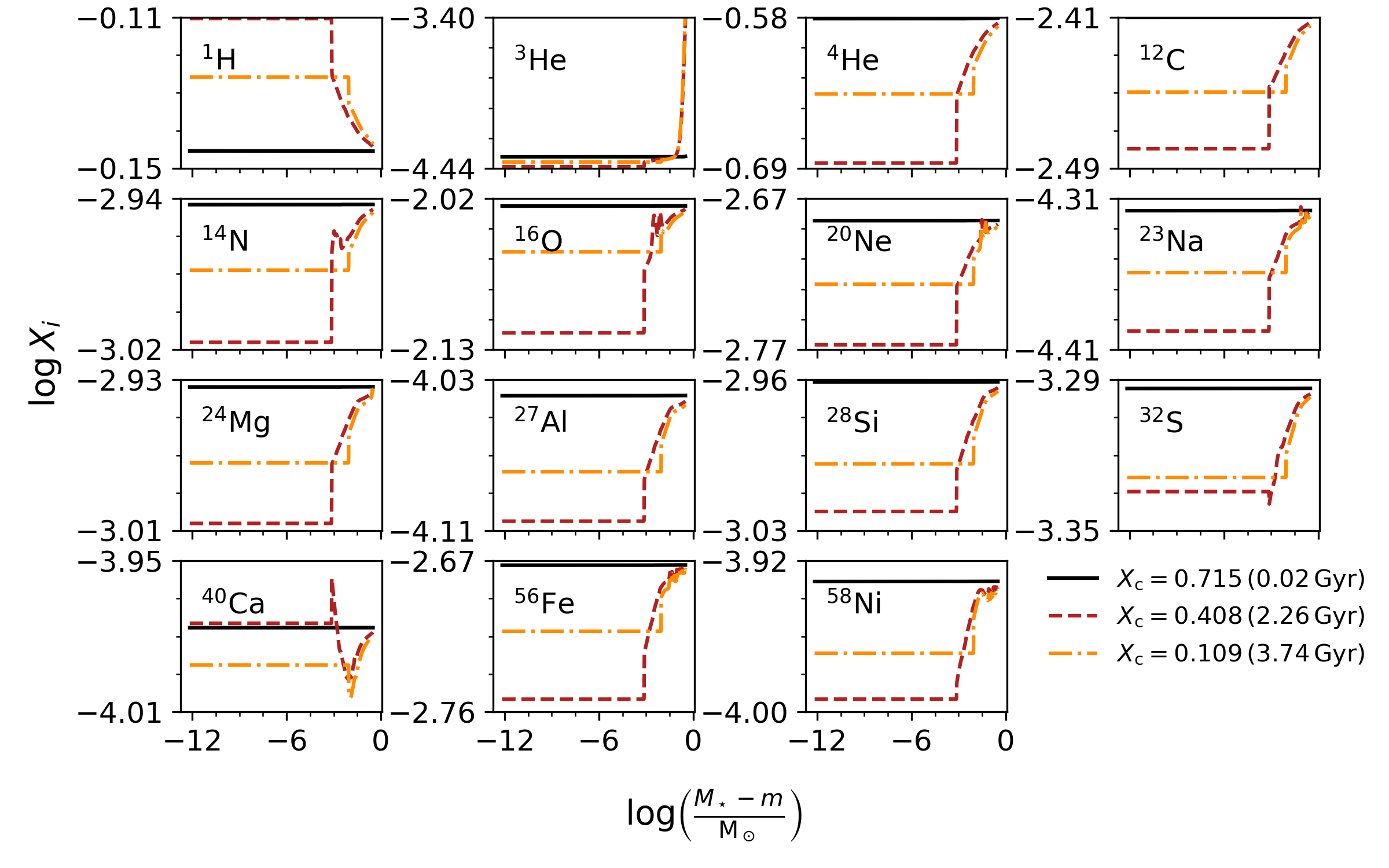}
    \caption{Evolution of the mass fractions throughout a 1.4\dMsun star with \zini = 0.022 and \fov = 0.0175 when atomic diffusion is active. The quantity $m$ indicates the mass enclosed within a sphere of radius $r$, hence the distance to the stellar center decreases when moving to the right along the abscissa.}
    \label{fig:massfraction}
\end{figure*}

\begin{figure}
    \centering
    \includegraphics[width = 0.48\textwidth]{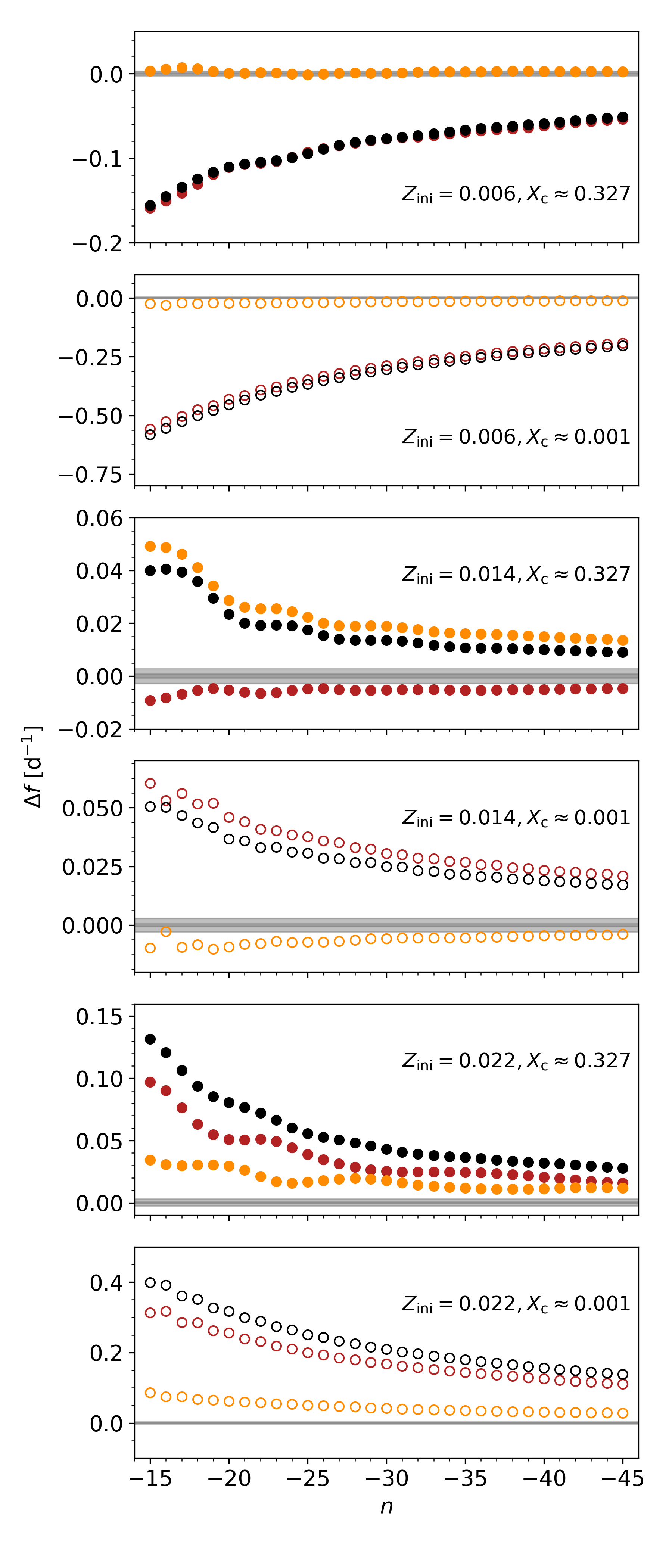}
    \caption{Difference in frequency ($\Delta f = f_{n, 1} - f_{n, 2}$, where model 2 is the baseline without diffusion) for a 1.4\dMsun model (\fov = 0.0175) as a function of radial order $n$. \textit{Black symbols:} baseline with standard OP tables. \textit{Orange symbols:} baseline with OP monochromatic opacities. \textit{Maroon symbols:} both without diffusion, but different opacity sources. The light and dark shaded areas mark the Rayleigh limit (1/$T$) for a 351-day (\textit{TESS} CVZ) and a 4-year timebase (\textit{Kepler} nominal mission), respectively. Rotation is not taken into account in the pulsation computations of this illustration.}
    \label{fig:df}
\end{figure}

The settling of heavier elements, mainly helium, has a stabilizing effect on the chemical gradient \citep{Theado2009} that results from a receding core as the local mean molecular weight is increased. Hence, the profile of the Brunt-V\"ais\"al\"a frequency is altered in the region where \gmodes are most sensitive. In Fig.~\ref{fig:effect_N+mu} the effect of diffusion on the chemical gradient ($\nabla_\mu$) and on the Brunt-V\"ais\"al\"a frequency ($N$) is demonstrated.
 The migration of the chemical elements for which \comm{radiative levitation is} computed is shown in Fig.~\ref{fig:massfraction} for three different moments throughout the main-sequence evolution.
Similar behavior is seen when we compare the behavior of the accumulation of the chemical elements shown in Fig.~\ref{fig:massfraction} to that found by \cite{Deal2018} (their Fig.~1, similar metallicity of $Z = 0.025$). The absence of a decrease in $X_i$ at the bottom of the surface convection zone for Fe and Al is \comm{the result of a different treatment of overshooting at the boundaries of internal convective shells and the bottom of the thin outer convective envelope.} \newline
Gravity-mode frequency shifts are the net result of the use of a different opacity source, and the changes in the chemical stratification introduced by atomic diffusion. Fig.~\ref{fig:df} shows the frequency differences between a model with atomic diffusion (OP monochromatic opacities), and two models without atomic diffusion, one with the OP monochromatic opacities, and one with the standard OP tables. The changes in input physics are mostly felt by the lower radial orders, as can be seen in Fig.~\ref{fig:df}. It should be noted that larger frequency differences occur for metallicities deviating from solar metallicity as a result of the uncertainties on the monochromatic opacity tables. As can be seen from Fig.~\ref{fig:df} the frequency shifts caused by atomic diffusion alone (orange points) are in any case larger than the Rayleigh limit of a 1-year long lightcurve. The frequency difference as a function of radial order seems to follow a trend, as the models without diffusion tend to either over- or underestimate the frequencies calculated from the models with diffusion. However, parametrizing a correction on the frequencies predicted by a model without atomic diffusion to obtain the predicted frequencies by a model with atomic diffusion is non-trivial as the frequency difference per radial order is dependent on the mass, age, and composition of the star, as is shown in Fig.~\ref{fig:df} for the latter two parameters. \newline

\section{Asteroseismic modeling} \label{sec:results}
\begin{figure}[htb]
    \centering
    \includegraphics[width = 0.48\textwidth]{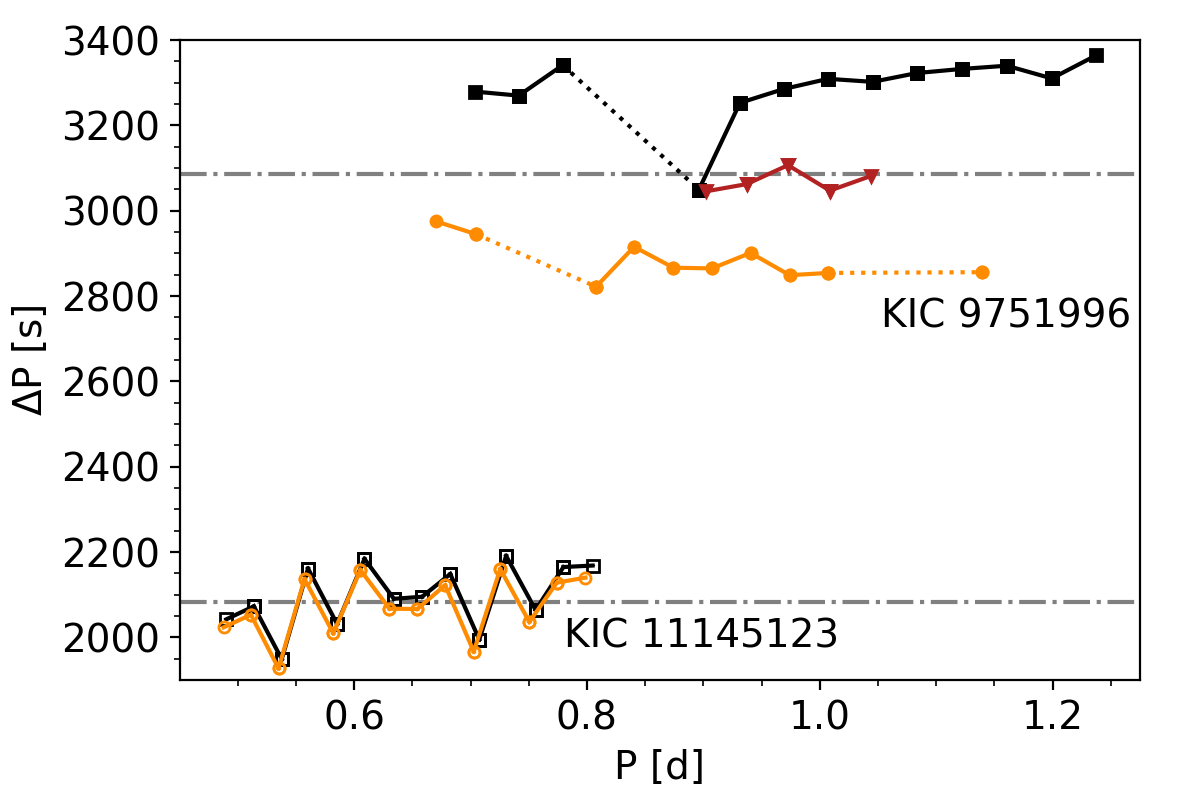}
    \caption{Period spacing patterns of \kn \citep{VanReeth2016} and \ko \citep{Kurtz2014}. The fitting is done as follows. For each azimuthal order (black squares: $m = -1$, maroon triangles: $m = 0$, orange circles: $m = 1$), the smallest period of each continuous sequence is assigned to a theoretical mode period and the consecutive periods to consecutive radial orders, starting with the longest sequence. The continuous sequences are separated by dotted lines. The gray lines indicate the measured values of $\Pi_0 / \sqrt{\ell (\ell + 1)}$, where $\ell = 1$.}
    \label{fig:p-dp}
\end{figure}

To date, the modeling of \gmodes in single \gDor stars, relied on \pin as asteroseismic input, combined with the \teff and \logg from spectroscopy \citep{Mombarg2019}. In this work, we investigate if replacing the asteroseismic input with the periods of the individual pulsations, instead of \pin, will improve upon the constraining of the mass and age, as this has never been studied for A/F-type g-mode pulsators. Commonly, the pulsations are presented in a so-called period spacing pattern, where the spacing between two periods of consecutive radial orders ($\Delta P = P_{n+1} - P_{n}$) is plotted as a function of $P_{n}$, which we have illustrated for \kn and \ko in Fig.~\ref{fig:p-dp}. The spectroscopic measurements have relatively larger uncertainties with respect to those for the periods of the \gmodes \citep[see Table~1 in][]{Aerts2019-ARAA}, and there are only two spectroscopic input parameters compared to about 30 periods for each of the two stars. Therefore, adding \teff and \logg will not contribute much to the goodness of fit, as we demonstrate in Sections~\ref{sec:fit_ND} and \ref{sec:fit_RL}. In addition, we also only fit the pulsations and use the spectroscopic measurements a posteriori to eliminate models that do not agree with the measured \teff and \logg within the 1$\sigma$ uncertainties. 
Selection of the best model is done by fitting each of the observed identified periods of the \gmodes for each azimuthal order. Starting with the longest period spacing sequence (see Fig.~\ref{fig:p-dp}), the lowest observed period, $P^{\rm obs}_{1}$, is assigned to the closest period of the theoretical model and the consecutive observed periods are then assigned to the consecutive radial orders. The radial order identification is then repeated for the next longest sequence (if present), demanding a radial order cannot be assigned multiple times. If this is the case, the model is removed from consideration.

 \cite{Aerts2018-apjs} describe the Mahalanobis distance as a merit function to account for the variance of the individual mode periods across a grid of stellar models to take into account correlations between and heteroscedasticity of the measured g-mode periods and the stellar parameters to be estimated. This merit function has already been applied to asteroseismic modeling using (\pin, \teff, \logg) by \cite{johnston2019} and \cite{Mombarg2019}, and using photometric colors in the context of clusters by \cite{Johnston2019-cluster}, but has never been applied to individual pulsations in \gDor stars. However, in this study we work with grids covering relatively small ranges \comm{for computational reasons and these do not allow to assess the variance-covariance matrix for the theoretically predicted pulsation periods in an adequate way.} For this reason, we use the simplified version of the Mahalanobis distance, which ignores the variance-covariance structure among the observed mode periods. This simplified version corresponds with the reduced $\chi^2$, defined as,
 \begin{equation}
     \chi^2_{\rm red} = \frac{1}{N - k}\sum_{i}^{N} \frac{\left(P^{\rm (th)}_i - P^{\rm (obs)}_i\right)^2}{\sigma_{P_i}^2},
 \end{equation}
 where $N$ is the number of observed g-mode periods and $k$ the number of free parameters.
 \newline
In addition to the pulsations, the fingerprints of atomic diffusion are also revealed at the stellar surface, as the process alters the surface abundances during the evolution of the star. Since the migration of each chemical species is different, from the chemical surface composition, one might be able to determine whether atomic diffusion is indeed active in a star, assuming an age and initial composition. For the models that best reproduce the observed periods and spectroscopic measurements, a comparison is made between the predicted and observed surface abundances, as an additional test for the influence of atomic diffusion. We stress that the surface abundances are not fitted, as both the observational and theoretical values have large uncertainties. 

Besides the surface abundances being altered by atomic diffusion, it might also be possible that the star was born with a solar metallicity, but its metal abundances at the surface have been altered as a result of atomic diffusion. Hence, both stars are also fitted to grids where we fix \zini~=~0.014 (same mass and overshoot ranges compared to the grids with the metallicity set according to the spectroscopic value). \comm{Since for solar metallicity the difference between the two discussed options of setting $Y_{\rm ini}$ is negligible, we only explore the option where $X_{\rm ini}$ is fixed for this case. }
The observed periods of both stars are fitted to the grids described in Table~\ref{tab:grids}, where we have used the OP monochromatic opacities for the diffusion models and the OP tables (non monochromatic) from \mesa for the models without diffusion. In order to make a meaningful comparison between the models with and without atomic diffusion, the Akaike Information Criterion (AIC) is evaluated. We use the AIC, corrected for a small sample size, defined as follows,
\begin{equation}
    {\rm AIC} =  \chi^2 + \frac{2kN}{N - k - 1},
\end{equation}
where $N$ is number of observed periods, and $k$ the number of free parameters. The grid with fixed metallicity has one degree of freedom less, while $k$ does not change when diffusion is taken into account. When comparing two models A and B, model B is favored over model A if $\Delta$AIC = AIC$_{\rm A}$ - AIC$_{\rm B}$ $\textgreater$ 2, where the evidence is (very) strong if $\Delta$AIC $\textgreater$ 6 (10) \citep{Kass1995}. These regimes stem from the value of $2\ln[P(D|{\rm A})/P(D|{\rm B})]$, where $P(D|{\rm X})$ is the probability of the observations $D$ given model X. For example, a difference $\Delta$AIC = 6 means the probability of producing the data with model A is roughly 95$\%$.  

\subsection{Models without atomic diffusion} \label{sec:fit_ND}
We first fit both stars to the respective grids without atomic diffusion listed in Table~\ref{tab:grids} to test if an adequate solution can be found, when the individual mode periods are used as input, instead of \pin as was done in \cite{Mombarg2019}. 

\subsubsection{\kn}
When using \teff and \logg from spectroscopy as additional input, along with the g-mode periods, we find a best model with the parameters listed in Table~\ref{tab:best_models_seis+spec_975} in Appendix \ref{app:other_fits} \comm{(only for fixed $X_{\rm ini}$)}. Given the observed values of \teff and \logg (Table~\ref{tab:GSSP}) and their respective uncertainties, the acquired model is consistent within $2\sigma$. When we demand the model which best fits the pulsations is consistent within the 1-$\sigma$ uncertainty intervals of these two spectroscopic observables, the model M01 listed in the first column of Table~\ref{tab:best_models} is found. \car{The choice of the initial composition impacts the obtained stellar parameters (see Table~\ref{tab:best_models}). This is expected as the predicted mode periods are degenerate with respect to mass and metal fraction, i.e. decreasing the mass of a model has the same effect on the mode periods as increasing the metal fraction \citep{Moravveji2015, Moravveji2016, Mombarg2019}. Similarly, increasing the metal fraction shifts \teff to cooler temperatures. Therefore, imposing a cutoff in \teff will yield more massive models.} \cite{Mombarg2019} estimated the mass of \kn to be $1.95 \pm 0.10$\Msun, situated on the MS around \cart{$X_{\rm c}/X_{\rm ini} = 0.20^{+0.10}_{-0.12}$ (where $X_{\rm ini} = 0.71$)}, assuming the same $D_{\rm CBM}(r)$ description as in Eq.~(\ref{eq:D_CBM}). However, that study assumed a metallicity close to the solar value. Indeed, when the observed [M/H] is assumed to be representative of the initial metallicity, a lower mass is found, as is expected from the well-known degeneracy between mass and metallicity for the predicted pulsation frequencies. 

\begin{figure}[htb]
    \centering
    \includegraphics[width = 0.49\textwidth]{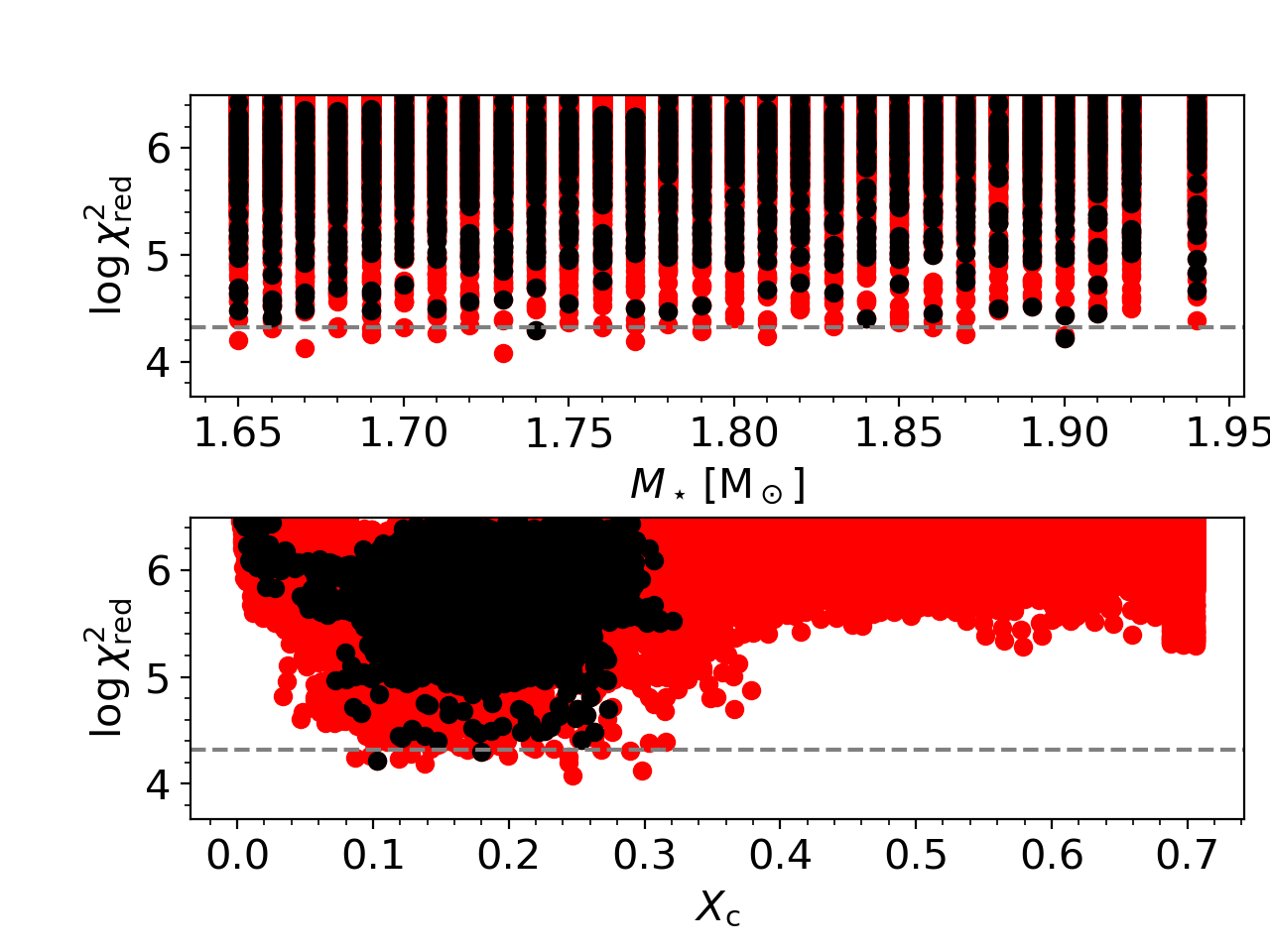}
    \caption{Reduced $\chi^2$ distributions when modeling \kn with the grid without diffusion where $Y_{\rm ini}$ is set as per \cite{Verma2019}. The red dots indicate the models which have been eliminated since these do not agree with the 1-$\sigma$ uncertainties on \teff and \logg. The dashed gray line indicates the 1-$\sigma$ confidence interval for the black dots. }
    \label{fig:theta-chi2_nodiff_975_EL}
\end{figure}

\begin{figure*}[htb]
    \centering
    \includegraphics[width = \linewidth]{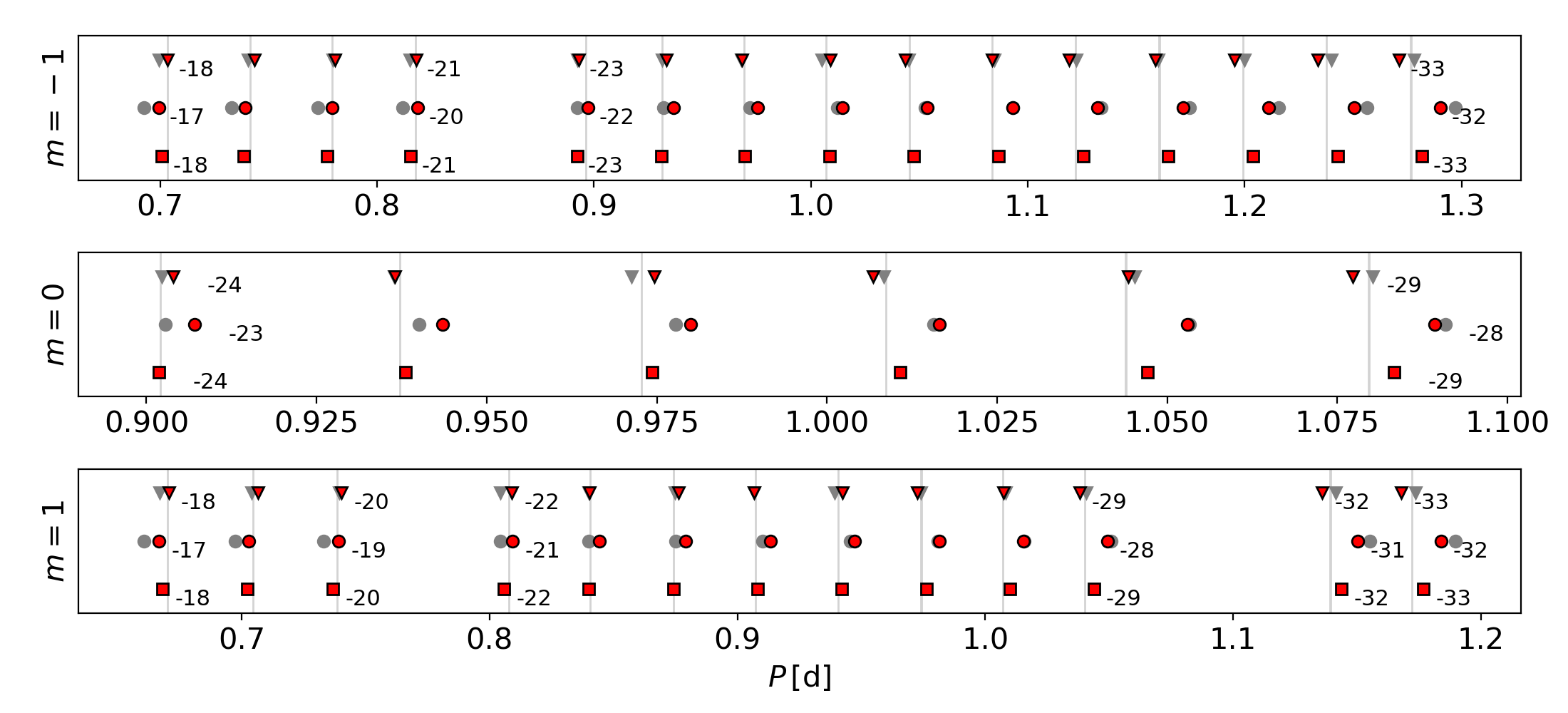}
    \caption{Observed mode periods of \kn indicated as gray vertical lines from \cite{VanReeth2015}, the uncertainties on the measured mode periods are too small to be seen at this scale. The symbols represent the theoretical predicted mode periods of the best-fitting models in each grid. \comm{Triangles: Model without diffusion and $Z_{\rm ini}$ according to the spectroscopically derived value (model M02). Circles: Model with diffusion and $Z_{\rm ini}$ according to the spectroscopically derived value (model M04). Squares: Model with diffusion and $Z_{\rm ini}$ fixed at 0.014 (model M05). Radial orders of the modes are indicated next to the symbols. The gray symbols indicate the mode periods of models M01 and M03.}}
    \label{fig:P_975_1s}
\end{figure*}

\begin{figure}[htb]
    \centering
    \includegraphics[width = 0.45\textwidth]{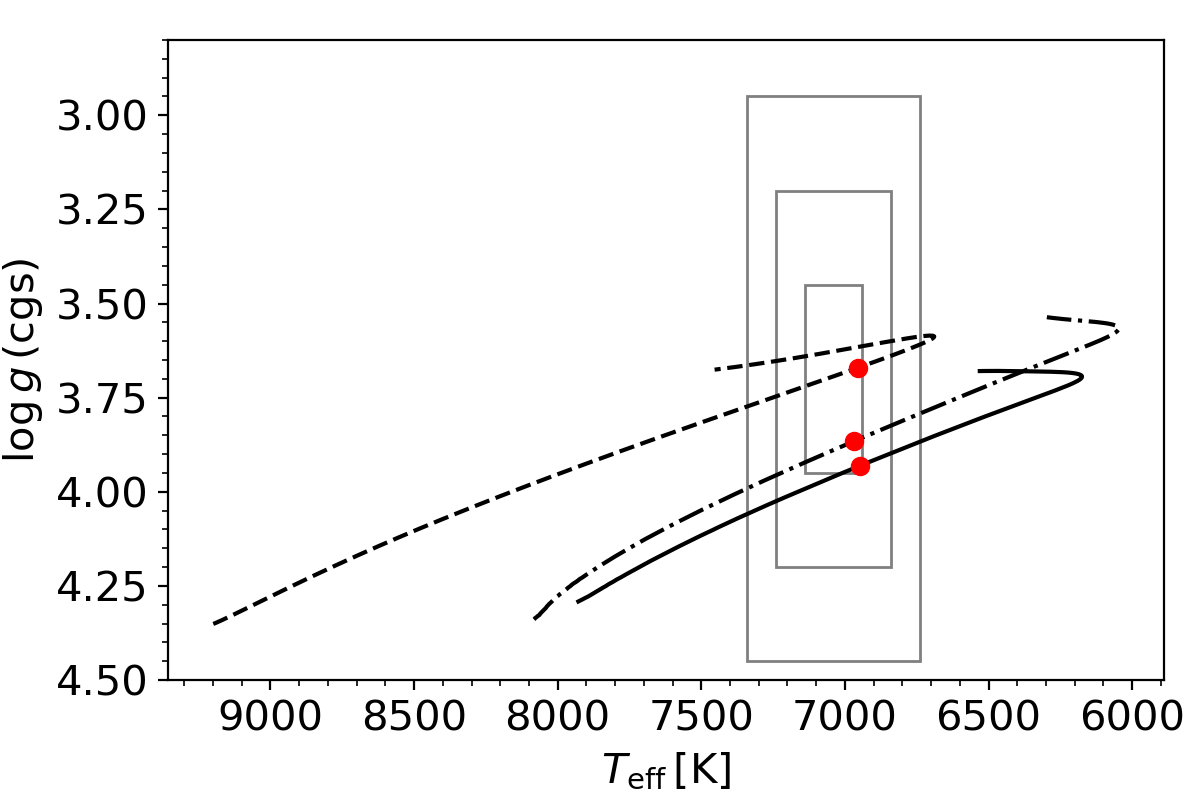}
    \caption{Evolution tracks for the best model with atomic diffusion (black solid line, model M04) and best model without diffusion (black dashed line, model M02) for \kn. The evolution track of the best model for with atomic diffusion and $Z_{\rm ini} = 0.014$ is indicated with the black dashed-dotted line (model M05). The red dots indicate the predicted \teff and \logg for each of the models. The gray boxes mark the 1-, 2-, and 3-$\sigma$ spectroscopic uncertainties.}
    \label{fig:975_HRD}
\end{figure}

As can be seen from the $\chi^2$ distributions in Fig.~\ref{fig:theta-chi2_nodiff_975_EL}, modeling the individual pulsation periods does not allow us to constrain the mass further in comparison to using (\pin, \teff, \logg) as input. \comm{For the two ways of setting the initial composition as discussed in Sec.~\ref{sec:methods}, we acquire a best solution. The solution with the lowest $\chi^2_{\rm red}$ out of these two is presented in Table~\ref{tab:grids} and the other solution is presented in Appendix~\ref{app:other_fits}. Setting the initial composition as per \cite{Verma2019}, does yield a slightly better fit compared to fixing $X_{\rm ini}$ (model M01; shown in gray in Fig.~\ref{fig:P_975_1s}).} \cart{We find that the estimated values for $M_\star$ and \xc are consistent with the values from \cite{Mombarg2019}.} 
The corresponding predicted periods of the best-fitting model after a 1-$\sigma$ cutoff in \teff and \logg are shown in the middle row of each panel in Fig.~\ref{fig:P_975_1s} \comm{(red/gray triangles; model M02/M01)} and the evolution track is shown in Fig.~\ref{fig:975_HRD}. The uncertainties of g-mode predictions from evolutionary models and pulsation computations are typically of order $10^{-3}$\,d$^{-1}$, which is two to three orders of magnitude above the uncertainties of observed modes from the nominal {\it Kepler\/} mission \citep[cf.][]{Aerts2018-apjs}. Hence, we are dealing with large $\chi^2$ as is well known from the modeling of g modes \citep[e.g.][]{Moravveji2016}.

\subsubsection{\ko}
\begin{figure}[htb!]
    \centering
    \includegraphics[width = 0.49\textwidth]{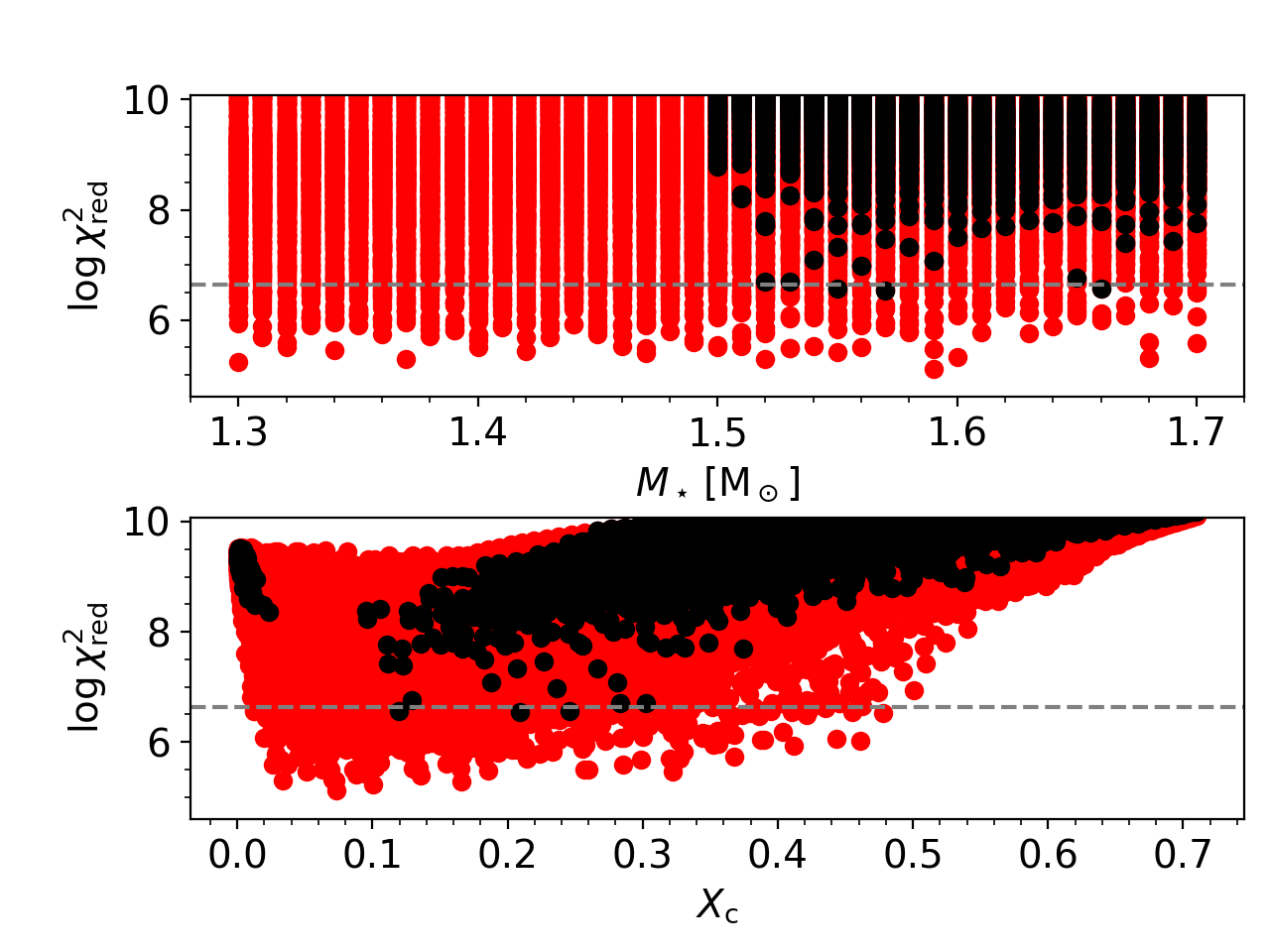}
    \caption{Reduced $\chi^2$ distributions when modeling \ko with the grid without diffusion where $Y_{\rm ini} = 1 - X_{\rm ini} - Z_{\rm ini}$. The red dots indicate the models which have been eliminated since these do not agree with the 1-$\sigma$ uncertainty on \teff and 3-$\sigma$ uncertainty on \logg. The dashed gray line indicates the 1-$\sigma$ confidence interval for the black dots. }
    \label{fig:theta-chi2_nodiff_111}
\end{figure}

Including \teff and \logg in the fit yields the best-fitting model listed in the first column of Table~\ref{tab:best_models_seis+spec_111} in Appendix~\ref{app:other_fits}. In this case, the predicted \teff and \logg are in disagreement with the observed values, and adding these two input parameters compared to 30 periods does not remedy this inconsistency. Moreover, the obtained mass is at the edge of the grid, below which we deem it is unlikely for a star to show g-mode pulsations. Eliminating models in disagreement with the spectroscopic values of \teff and \logg instead forces the best solution to the higher edge of the mass coverage. Therefore, we have expanded the grids without atomic diffusion from 1.5\Msun to 1.7\Msun (cf. Fig.~\ref{fig:theta-chi2_nodiff_111}).

\begin{figure*}
    \centering
    \includegraphics[width = \linewidth]{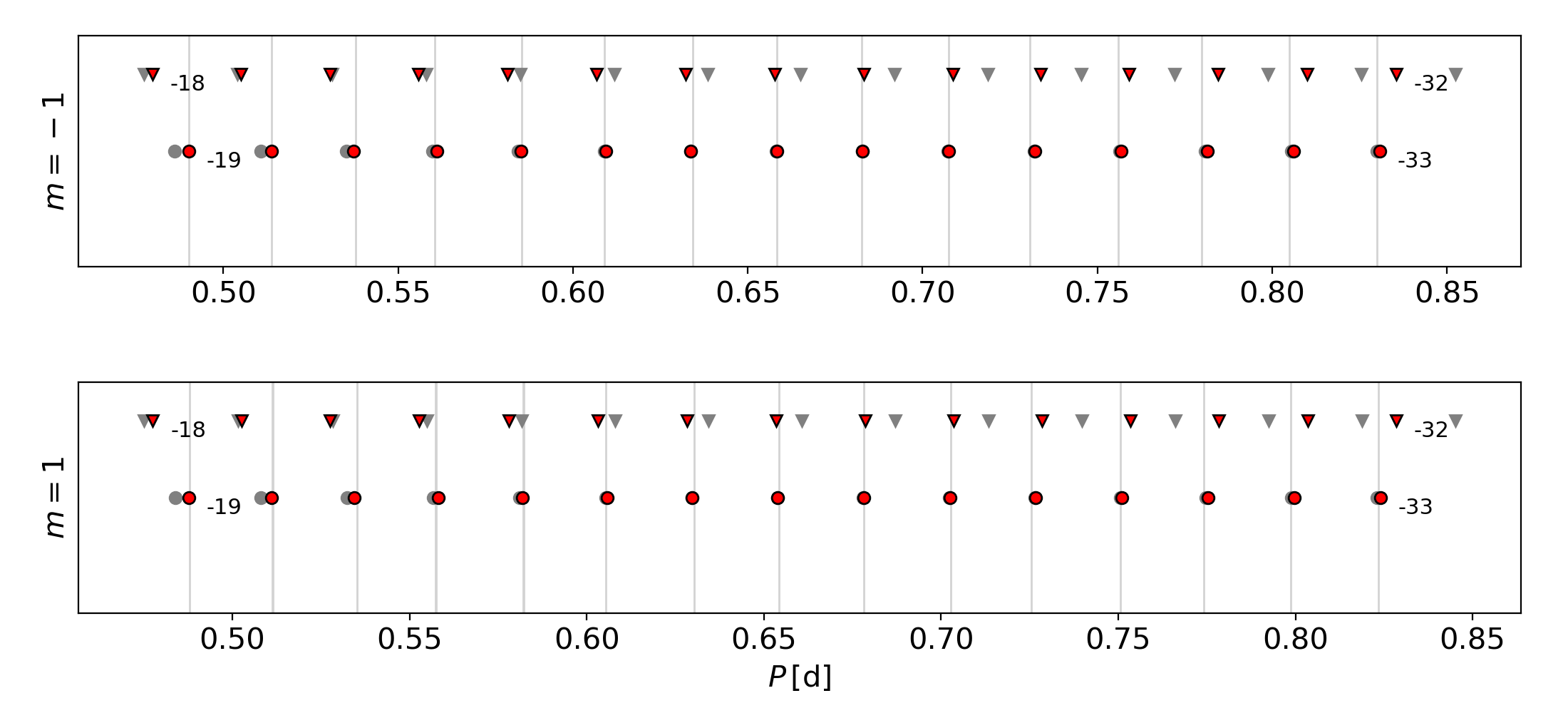}
    \caption{Observed mode periods of \ko indicated as gray vertical lines from \cite{Kurtz2014}, the uncertainties on the measured mode periods are too small to be seen at this scale. The symbols represent the theoretical predicted mode periods of the best-fitting models in each grid. \comm{Triangles: Model without diffusion and $Z_{\rm ini}$ according to the spectroscopically derived value according to \cite{Takada-Hidai2017} (model M06). Circles: Model with diffusion and $Z_{\rm ini}$ according to the spectroscopically derived value (model M08). Radial orders of the modes are indicated next to the symbols. The gray symbols indicate the mode periods of models M07 and M09.}}
    \label{fig:P_111_1+3s}
\end{figure*}

\begin{figure}
    \centering
    \includegraphics[width = 0.45\textwidth]{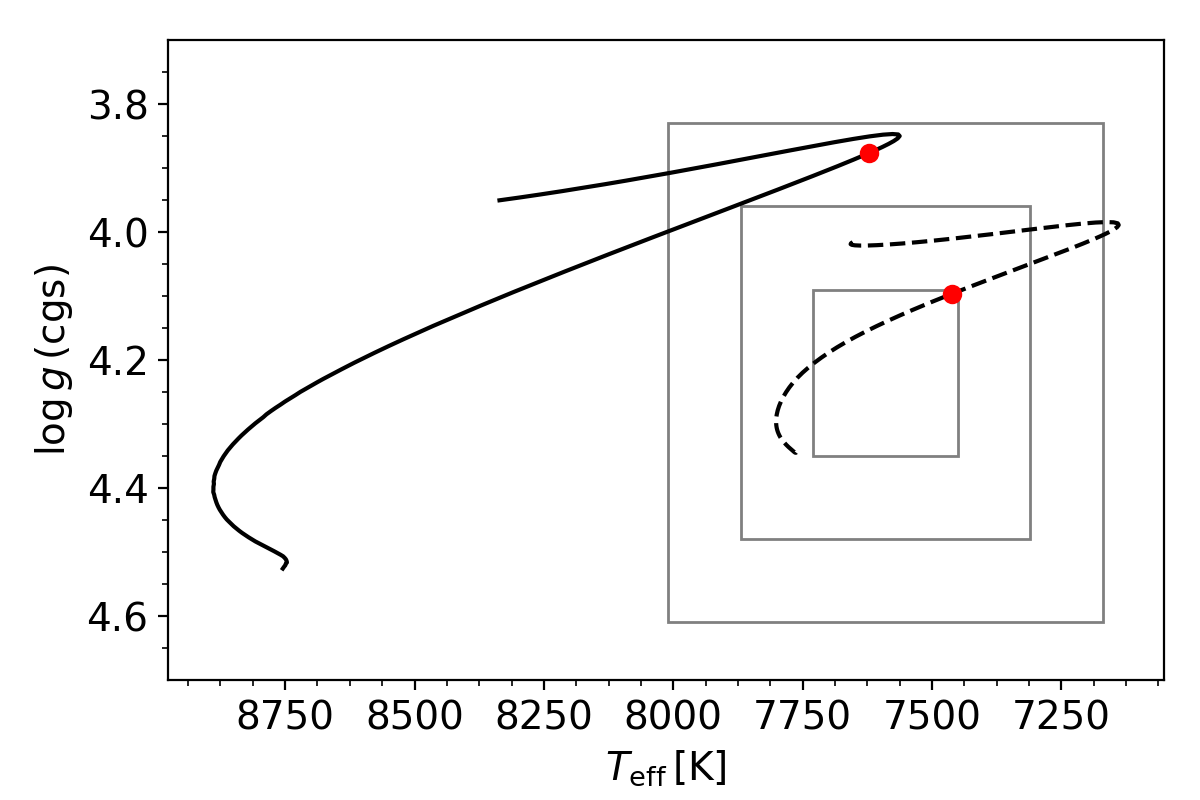}
    \caption{Same as Fig.~\ref{fig:975_HRD}, but for \ko. The spectroscopic \teff and \logg are taken from \cite{Takada-Hidai2017}, where the more conservative uncertainty on \teff (Table~\ref{tab:GSSP}) is taken.}
    \label{fig:111_HRD}
\end{figure}
\ko is most likely, judging from the value of \pin and the low rotation rate, near the end of hydrogen-core burning, as is found by \cite{Kurtz2014}, who did not have any spectroscopic measurements for the star at the time of their asteroseismic interpretation. However, the measured \logg~$= 4.22 \pm 0.13$ from \cite{Takada-Hidai2017} would suggest the star is in the earlier phase of the MS. Therefore, we relax the demand on \logg and allow it to be within 3$\sigma$ instead, yielding the model listed in Table~\ref{tab:best_models} (see Table~\ref{tab:best_models_seis+spec_111} in the appendix for the 1-$\sigma$ cutoff solution). In case of \ko, we are not able to fit the observed mode periods as well as for \kn when atomic diffusion is not taken into account as is shown in Fig.~\ref{fig:P_111_1+3s} (triangles). The evolution track of the best-fitting model \comm{(M06)} is shown in Fig.~\ref{fig:111_HRD}. \comm{We find that when a higher $Y_{\rm ini}$ is assumed, the observed mode periods are better matched, compared to when $Y_{\rm ini}$ is set to the predicted value according to the enrichment rate by \cite{Verma2019} (models M07 and M09; gray symbols in Fig.~\ref{fig:P_111_1+3s}). }

\begin{table*}[]
    \centering
    \begin{tabular}{l|lllll|lllll}
    \hline \hline
    \multicolumn{1}{c}{} & \multicolumn{5}{c}{\kn} & \multicolumn{5}{c}{\ko} \\
    \hline
    Model ID & M01 & M02 & M03 & M04 & M05 & M06 & M07 & M08 & M09 & M10 \\
Diffusion                        & No    &  {No}      & Yes     &  {Yes}    &  {Yes}     &  {No}     & No      &  {Yes}    & Yes   & Yes \\
$M_\star\,[{\rm M_\odot}]$      & 1.71   &  {1.90}    & 1.89   &  {1.80}    &  {1.68}    &  {1.57}   & 1.66    &  {1.36}   & 1.43      & - \\
$X_{\rm c}$                     & 0.256  &  {0.103}   & 0.292  &  {0.314}   &  {0.286}   &  {0.209}  & 0.292   &  {0.058}  & 0.095      & - \\
$X_{\rm ini}$                  & 0.715  &  {0.698}   & 0.715  &  {0.707}   &  {0.715}   &  {0.715}  & 0.752   &  {0.715}  & 0.749      & 0.715 \\
$Y_{\rm ini}$                   & 0.259  &  {0.276}   & 0.263  &  {0.271}   &  {0.271} &  {0.283}  & 0.246   &  {0.283}  & 0.248      & 0.271 \\
$Z_{\rm ini}$                   & 0.026  &  {0.026}   & 0.022  &  {0.022}   &  {0.014}   &  {0.002}  & 0.002   &  {0.002}  & 0.003      & 0.014 \\
$f_{\rm ov}$                    & 0.0100 &  {0.0175}  & 0.0100 &  {0.0175}  &  {0.0300}  &  {0.0100} & 0.0100  &  {0.0225} & 0.0100      & - \\
$\log \chi^2_{\rm red}$         & 4.23   &  {4.22}    & 5.45   &  {5.18}    &  {4.45}    &  {6.54}   & 7.79    &  {5.34}   & 5.50      & - \\
$\tau\,$[Gyr]                   & 1.290  &  {1.139}   & 1.116  &  {1.297}   &  {1.689}   &  {1.484}  & 1.402   &  {2.234}  & 2.075      & - \\
$M_{\rm cc}\,[{\rm M_\odot}]$   & 0.177  &  {0.169}   & 0.190  &  {0.189}   &  {0.193}   &  {0.100}  & 0.103   &  {0.132}  & 0.121      & - \\
$T_{\rm eff}\,$[K]              & 7137   &  {6953}    & 7110   &  {6948}    &  {6968}    &  {7461}   & 7562    &  {7622}   & 7574      & - \\
$\log g\,$(cgs)                 & 3.95   &  {3.67}    & 3.94   &  {3.93}    &  {3.87}    &  {4.10}   & 4.13    &  {3.88}   & 4.02       & - \\
$\Pi_0\,$[s]                    & 4438   &  {4438}    & 4639   &  {4653}    &  {4475}    &  {3086}   & 3248    &  {2999}   & 2972      & - \\
$\Pi_0^{\rm (obs)}\,$[s] & \multicolumn{5}{c|}{$4364 \pm 7$} & \multicolumn{5}{c}{$2945 \pm 78$} \\
    \hline

    \end{tabular}
    \caption{Best-fitting parameters for both \kn and \ko when only models within 1$\sigma$ (3$\sigma$ for \logg in case of \ko) of the spectroscopically derived values of \teff and \logg are considered. All parameters listed below the reduced $\chi^2$ follow from the model and are thus not free parameters. \comm{When the metallicity is set according to the spectroscopic value, we list both solutions for the two different relations to set $X_{\rm ini}$ and $Y_{\rm ini}$, as discussed in Sec.~\ref{sec:methods}.} For \ko, the grid with fixed $Z_{\rm ini} = 0.014$ does not yield solutions consistent with the spectroscopic \teff and \logg. The bottom rows lists the measured values of \pin from \cite{VanReeth2016} and \cite{Kurtz2014} for \kn and \ko, respectively.}
    \label{tab:best_models}
\end{table*}

\subsection{Models with atomic diffusion} \label{sec:fit_RL}
We now fit both stars to the respective grids with atomic diffusion included (same parameter ranges) to test if the theoretical predictions yield a better match with the observations. As mentioned before, when atomic diffusion is included, the measured metallicity at the surface may not be representative of the initial metallicity. Therefore, we also fit both stars to grids where we fix the initial metallicity, $Z_{\rm ini} = 0.014$. 
\subsubsection{\kn} 
\begin{figure}[htb]
    \centering
    \includegraphics[width = 0.49\textwidth]{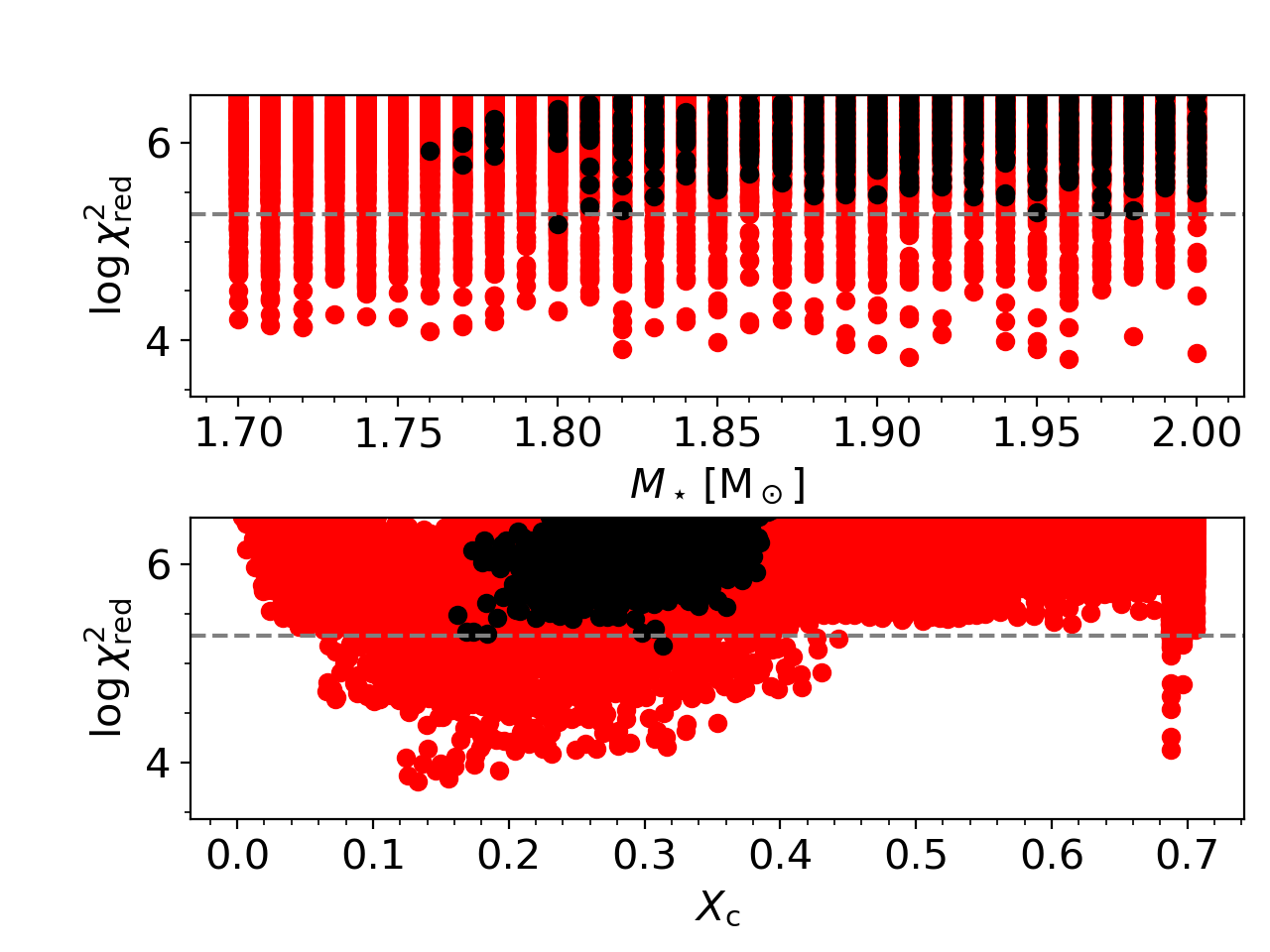}
    \caption{Same as Fig.~\ref{fig:theta-chi2_nodiff_975_EL} (\kn), but for the grid with atomic diffusion included and a metallicity according to the spectroscopically derived value.}
    \label{fig:theta_chi2_radlev_975_EL}
\end{figure}

\begin{figure}[htb]
    \centering
    \includegraphics[width = 0.49\textwidth]{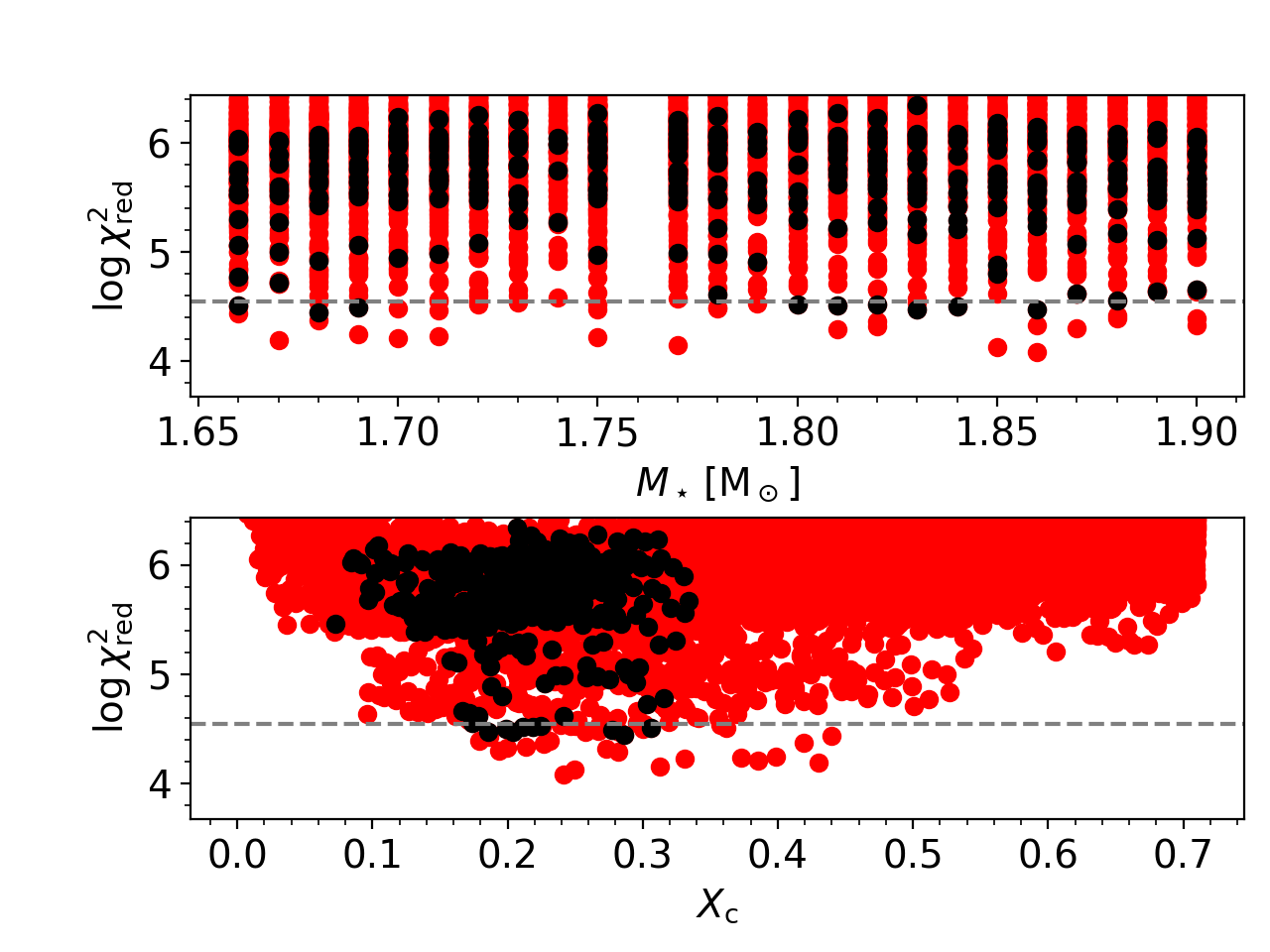}
    \caption{Same as Fig.~\ref{fig:theta-chi2_nodiff_975} (\kn), but for the grid with atomic diffusion included and a metallicity fixed at $Z_{\rm ini} = 0.014$.}
    \label{fig:theta_chi2_radlev_975_Z140}
\end{figure}

For the grids with atomic diffusion, we also explored both methods of including the spectroscopic parameters in the fit or using these parameters as a posteriori elimination criterion. The results of the primer method are presented in Table~\ref{tab:best_models_seis+spec_975} in Appendix~\ref{app:other_fits}, for which we again find that adding \teff and \logg does not enforce solutions consistent with spectroscopy. Table~\ref{tab:best_models} summarizes the best-fitting models when we apply the 1$\sigma$ cutoff in \teff and \logg. The corresponding $\chi^2$ distributions are shown in Figs~\ref{fig:theta_chi2_radlev_975_EL} and \ref{fig:theta_chi2_radlev_975_Z140}. When the metallicity is taken according to the spectroscopically determined value, we find a higher mass compared to the solution where atomic diffusion is not included. The mass-metallicity relation is still seen in the solutions when \teff and \logg are also fitted \comm{(for fixing $X_{\rm ini}$ only)}, but this improves when we eliminate the models inconsistent with the measured spectroscopy. All grids with atomic diffusion included give \xc estimates \comm{(models M03, M04, M05)} which are \cart{higher than} the estimated value by \cite{Mombarg2019}. The determined $\Pi_0 = 4364 \pm 7\,{\rm s}$ of \kn by \cite{VanReeth2015} cannot be reproduced by any of the best models, with or without atomic diffusion, and regardless whether \teff and \logg are fitted or used as a cutoff. 

\begin{figure}[htb]
    \centering
    \includegraphics[width = 0.40\textwidth]{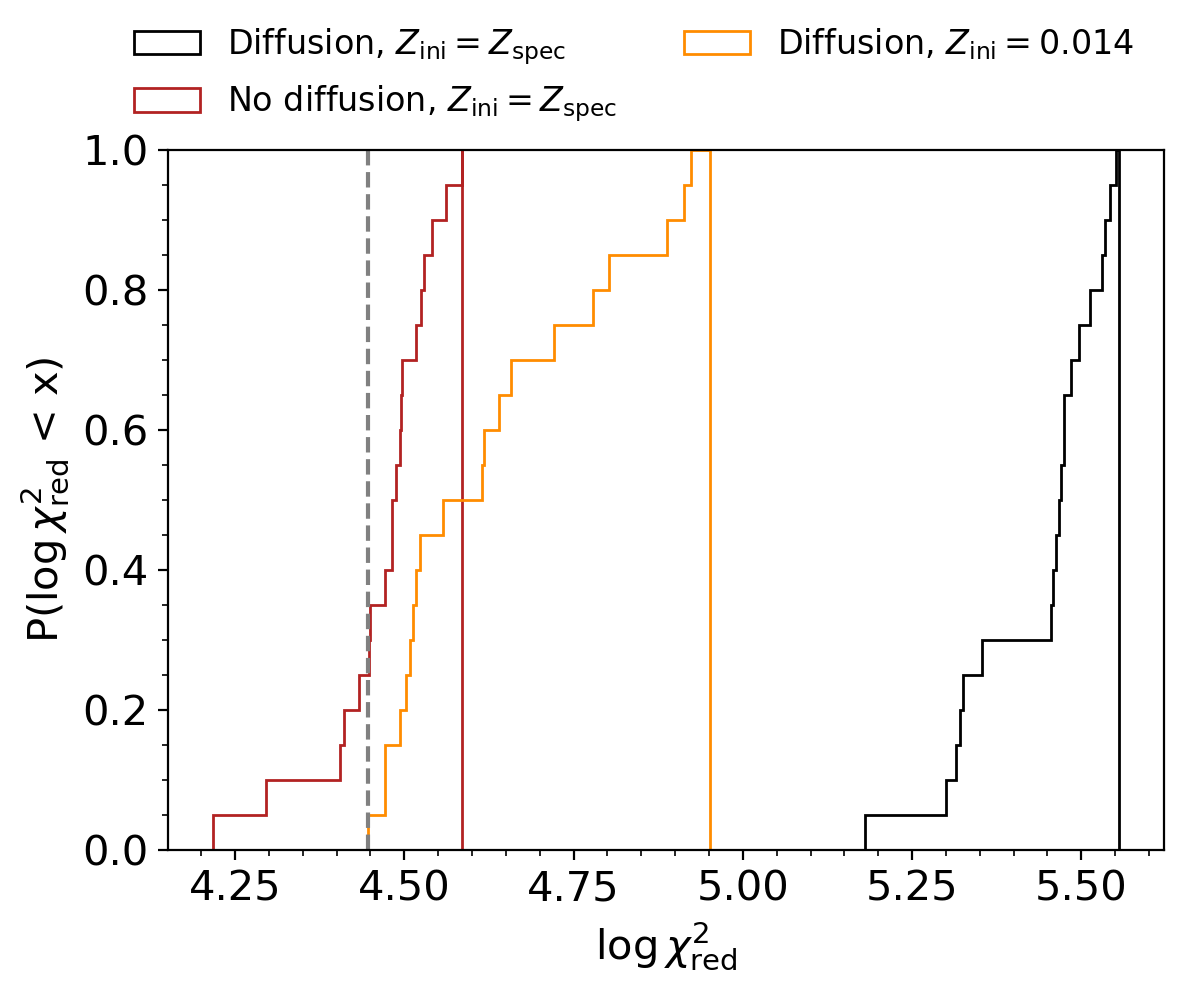}
    \caption{The $\log \chi^2_{\rm red}$ cumulative distributions of the best 20 models from each grid \comm{(corresponding to models M02, M04, and M05)} for \kn. As a visual aid, a dashed line is plotted at the $\log \chi^2_{\rm red}$ value of the best model from the grid where atomic diffusion is included, and the metallicity fixed at $Z_{\rm ini} = 0.014$. }
    \label{fig:hist_975}
\end{figure}

\begin{figure*}[htb]
    \centering
    \includegraphics[width = 0.9\linewidth]{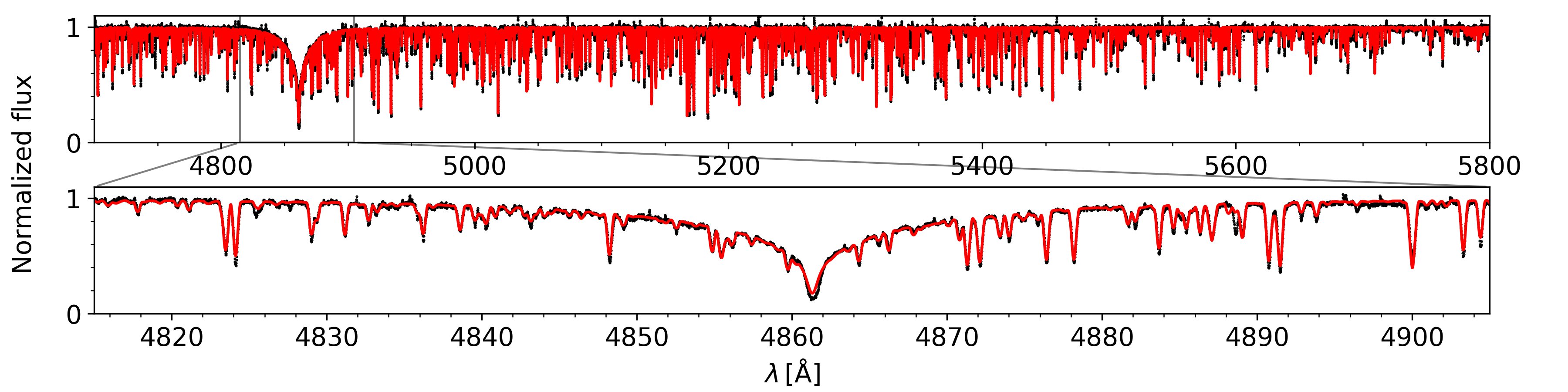}
    \caption{The best-fitting synthetic spectrum (red) for the normalized averaged spectrum of \kn (black). The bottom panel shows a zoom-in around the H$\beta$ line. }
    \label{fig:synth_spec}
\end{figure*}

The corresponding periods and evolution tracks of the best models \comm{(M04 and M05)} with atomic diffusion (spectroscopic and solar metallicity) are shown in Fig.~\ref{fig:P_975_1s} and Fig.~\ref{fig:975_HRD}, respectively. \comm{Fig.~\ref{fig:P_975_1s} also shows the predicted mode periods from model M03 (gray circles).}
Based on the difference in the value of the AIC, $\Delta$AIC, the models without atomic diffusion is strongly favored over the models without diffusion. \comm{In particular, for both choices of initial composition a model without atomic diffusion gives a better result. } In Fig.~\ref{fig:hist_975}, we show the cumulative \chir distributions of the three grids for the best 20 models. It can be seen that the grid with atomic diffusion with the metallicity set according to the spectroscopic value does not give a more accurate fit. \comm{About a quarter of the best models without atomic diffusion give a better match than the best models with atomic diffusion and $Z_{\rm ini} = 0.014$.}

Additionally, we study the star's surface abundances and compare these with the predictions from our best models. \comm{There are no abundance measurements available for KIC 9751996 available in the literature. Moreover, the spectra} of \kn used in the study by \cite{VanReeth2016} do not have the required signal-to-noise ratio (SNR) to derive surface abundances. Therefore, four additional spectra -- each with an exposure time of \SI{2700}{\sec} -- were taken on 2019 May 24-25 with the High Efficiency and high Resolution Mercator \'Echelle Spectrograph \citep[HERMES,][]{Raskin2011} on the 1.2-m Mercator telescope (La Palma, Spain). The spectra were taken with the HERMES high-resolution fiber, yielding a resolution of $R = \SI{85000}{}$. After normalization, the average spectrum is calculated. This normalized spectrum was fitted with synthetic spectra deduced from atmosphere models.
First, a global solution was found for the effective temperature \teff, surface gravity \logg, projected surface velocity \vsini, metallicity [M/H], and microturbulence $\xi$, using the \texttt{GSSP} software package \citep{Tkachenko2015}, where we use the spectrum between 4700 and 5800\,\AA. The parameters of the global solution and the measured surface abundances are presented in Table~\ref{tab:GSSP} and a plot of the best fitting synthetic spectrum is shown in Fig.~\ref{fig:synth_spec}. Our atmospheric parameters for the global solution are consistent within 1$\sigma$ compared to those by \cite{VanReeth2015}. Subsequently, the parameters of the global solution are fixed, and individual surface abundances are determined. The results are presented in Table~\ref{tab:surf_abun} which have been used as a check after the best asteroseismic models were selected. 

\begin{figure*}[htb]
    \centering
    \includegraphics[width = \linewidth]{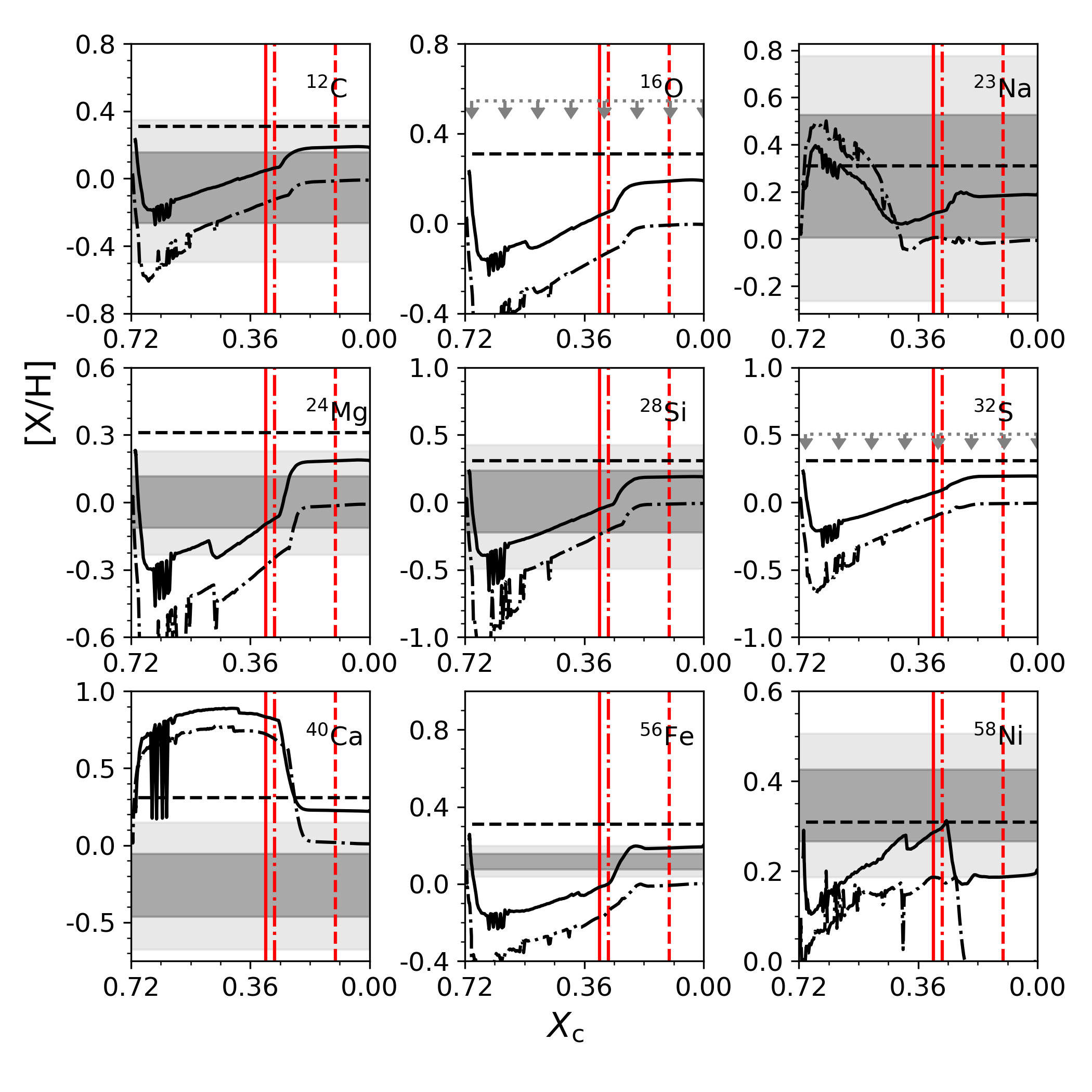}
    \caption{The evolution of the surface abundances of the best models M02, M04, and M05 for \kn. \textit{Solid line:} Model with atomic diffusion and \zini according to the spectroscopic value from \cite{VanReeth2016}. \textit{Dashed line:} Model without diffusion and \zini according to the spectroscopic value. \textit{Dashed-dotted line:} Model with diffusion and \zini = 0.014. The 1(2)-$\sigma$ intervals are shown as dark (light) shaded regions. For O and S, only the upper limits could be inferred. The red lines indicate the estimate of \xc for each model, where the same line style convention is used.}
    \label{fig:surf_abun_975}
\end{figure*}

Fig.~\ref{fig:surf_abun_975} shows the predicted surface abundances and the observed abundances of elements for which radiative \comm{levitations have} been computed. We refer to Appendix~\ref{app:conv} for the conversion of the abundances from $\log(n_{\rm X}/n_{\rm tot})$ to [X/H].
Regarding the theoretically predicted abundances, we point out that the sharp variations seen in some of the elements have a numerical origin, as at these points \grad is of the same order as $g$. This behavior is one of the reasons why we refrain from including the surface abundances in the selection scheme, as only the global trend should be trusted. Since the constant chemical mixing in the stellar envelope is at a low level to keep compliance with the g-mode trapping properties \citep{VanReeth2016}, the surface abundances of the models without atomic diffusion will remain constant throughout the MS evolution. The best model without atomic diffusion (solar mixture as per \cite{Asplund2009}) is not able to explain all observed surface abundances within the 2-$\sigma$ intervals. Yet, the models with atomic diffusion do not give more accurate predictions. Hence, the surface abundances do not help to improve the asteroseismic best model fit for in the case of \kn.

\begin{table}[]
    \centering
    \begin{tabular}{lr@{$\,\pm\,$}lr@{$\,\pm\,$}ll}
      & \multicolumn{2}{l}{Value} & \multicolumn{2}{l}{Value}& Closest \\
     Parameter &\multicolumn{2}{l}{\ko}  & \multicolumn{2}{l}{\kn}     & grid point \\
\hline
        
        $T_{\rm eff}~$[K] & \multicolumn{2}{c}{$7590^{+80}_{-140}$}  & 7040 & 100 & 7100 \\
        $\log g~$(cgs) & 4.22& 0.13 & 3.70 & 0.25 & 3.70 \\
        $v \sin i~$[${\rm km~s}^{-1}$] & 5.9 & 0.2$^\dagger$& 12.5 & 0.7 & 12 \\
        $\rm [M/H]$ &-0.71 & 0.11 & 0.20 & 0.07 & 0.2 \\
        $\xi \,$[{\rm km~s}$^{-1}$] &3.1 & 0.5 &  3.16 & 0.30 & 3.3 \\
    \end{tabular}
    \caption{Parameters for the best spectroscopic model of \ko according to \cite{Takada-Hidai2017} and \kn from this work using \texttt{GSSP}. The fourth column lists the value of the grid point closest to the best solution listed in the previous column. The listed [M/H] values are scaled  by assuming $Z_\odot = 0.0134$ as per \cite{Asplund2009}. $^\dagger$Apparent projected rotation velocity.  }
    \label{tab:GSSP}
\end{table}

\begin{table}[]
    \centering
    \begin{tabular}{lc}
    Element & $\log\left(n_{\rm X}/n_{\rm tot}\right)$ \\
        \hline
        
        C  & ${-3.64}^{+0.19}_{-0.23}$ \\
        O  & \textless${-2.8}$ \\
        Na & ${-5.52}^{+0.25}_{-0.27}$ \\
        Mg & ${-4.43}^{+0.11}_{-0.12}$ \\
        Si & ${-4.48}^{+0.19}_{-0.27}$ \\
        S  & \textless${ -4.41}$ \\
        Ca & ${-5.95}^{+0.20}_{-0.21}$ \\
        Sc & ${-9.22}^{+0.29}_{-0.35}$ \\
        Ti & ${-7.06}^{+0.11}_{-0.11}$ \\
        Cr & ${-6.06}^{+0.08}_{-0.09}$ \\
        Mn & ${-6.44}^{+0.21}_{-0.23}$ \\        
        Fe & ${-4.42}^{+0.04}_{-0.04}$ \\
        Y  & ${-9.68}^{+0.22}_{-0.25}$ \\
        Ni & ${-5.47}^{+0.08}_{-0.08}$ \\
    \end{tabular}
    \caption{Measured surface abundances, in number density per total number density, of \kn by fixing the parameters of the global solution listed in Table~\ref{tab:GSSP}.}
    \label{tab:surf_abun}
\end{table}

\subsubsection{\ko}
\begin{figure}[htb]
    \centering
    \includegraphics[width = 0.49\textwidth]{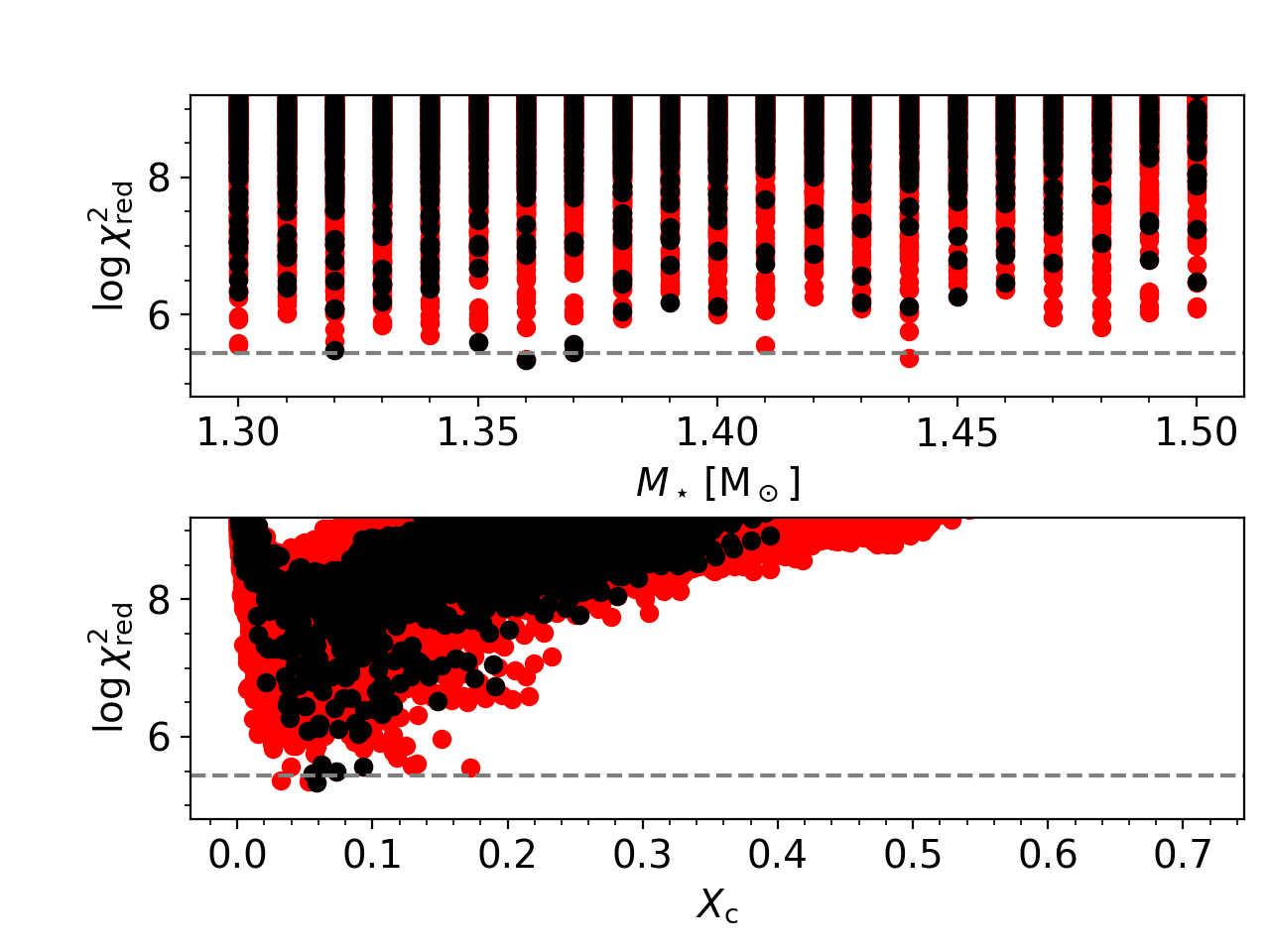}
    \caption{Same as Fig.~\ref{fig:theta-chi2_nodiff_111} (\ko), but for the grid with atomic diffusion included and a metallicity according to the spectroscopically derived value.}
    \label{fig:theta_chi2_radlev_111}
\end{figure}

\begin{figure}[htb]
    \centering
    \includegraphics[width = 0.49\textwidth]{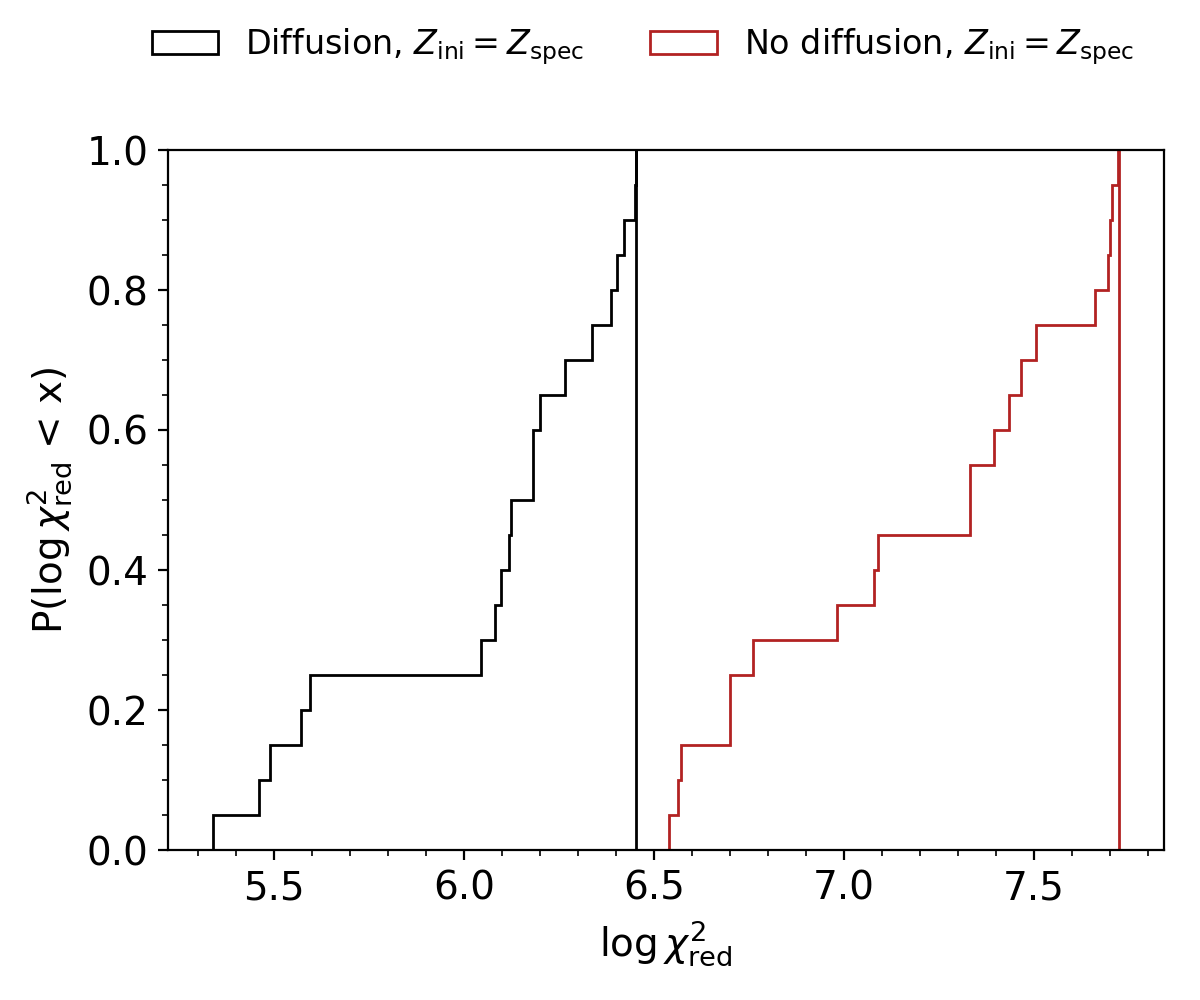}
    \caption{The $\log \chi^2_{\rm red}$ probability distributions of the best 20 models from each grid \comm{(corresponding to models M06 and M08)} for \ko (no consistent models found for the grid with solar metallicity).}
    \label{fig:hist_111}
\end{figure}

\begin{figure*}
    \centering
    \includegraphics[width = \linewidth]{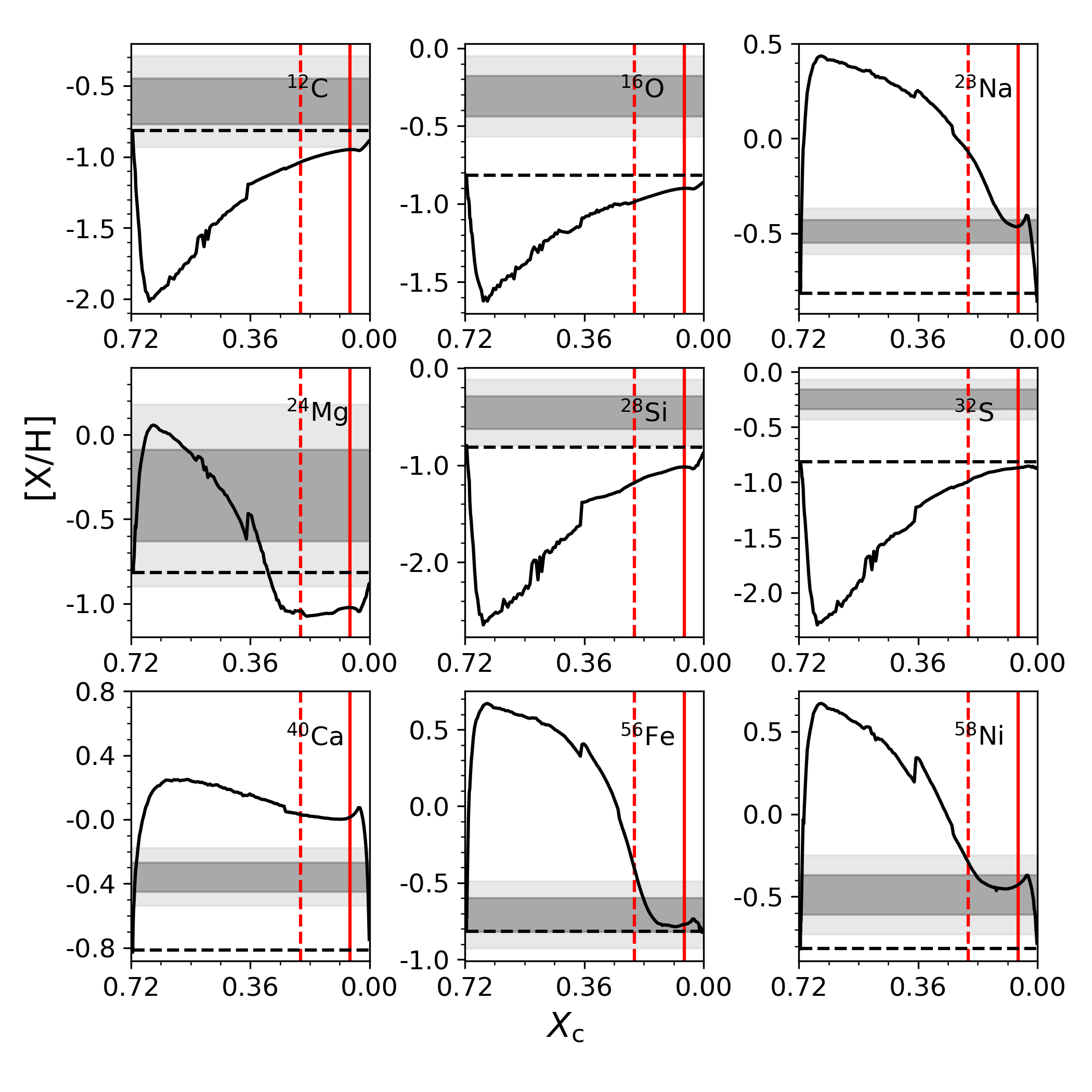}
    \caption{The evolution of the surface abundances of the best models M06 and M08 (no solution for $Z_{\rm ini} = 0.014$) for \ko. \textit{Solid line:} Model with atomic diffusion and \zini according to the spectroscopic value from \cite{Takada-Hidai2017}. \textit{Dashed line:} Model without diffusion and \zini according to the spectroscopic value. The 1(2)-$\sigma$ intervals from \cite{Takada-Hidai2017} are shown as dark (light) shaded regions. The red lines indicate the estimate of \xc for each model, where the same line style convention is used.   }
    \label{fig:surf_abun_111}
\end{figure*}

When \teff and \logg are taken as additional input parameters, along with the g-mode periods, we find that the model with the spectroscopic metallicity is consistent with the 1- and 3-$\sigma$ uncertainty intervals of \teff and \logg, respectively. However, the best model obtained for $Z_{\rm ini} = 0.014$ is inconsistent with these intervals. Table~\ref{tab:best_models} lists the best models when we demand compliance with the spectroscopy. The grid with \zini according to the spectroscopic surface metallicity returns a model that is near the TAMS, as predicted by \cite{Kurtz2014} (cf. Fig.~\ref{fig:theta_chi2_radlev_111}). The corresponding model pulsation periods are shown in Fig~\ref{fig:P_111_1+3s}, and the evolution track is shown in Fig.~\ref{fig:111_HRD}. If $Z_{\rm ini} = 0.014$ is assumed, no models can be found within the aforementioned spectroscopic intervals. We find that the grid with atomic diffusion gives a better fit compared to grid without, and is significantly more probable of reproducing the data according to the AIC. From Fig.~\ref{fig:hist_111}, it can be seen that all of the 20 best models with atomic diffusion give smaller $\chi^2$ values than the best model without atomic diffusion. \comm{The conclusion on whether atomic diffusion improves the modeling of the oscillations is independent of the two choices of the initial composition.}
The predicted surface abundances of the model with atomic diffusion is able to explain the Na and Ni surface abundances, whereas the model without atomic diffusion cannot, as shown in Fig.~\ref{fig:surf_abun_111}. The observed abundances of O, Si, S, and Ca are not reproduced by either model, which might be the result of the star having a different chemical mixture than the Sun.  

\newpage
\section{Conclusion \& Discussion} \label{sec:conclusion}
This work sheds light on four important questions regarding the asteroseismic modeling of slowly rotating $\gamma$ Doradus (A/F-type) pulsators. We investigated the effect of including the process of atomic diffusion, with \comm{accelerations induced by radiative levitation} included, in the equilibrium models for intermediate-mass stars, and the consequence for the frequencies of the \gmodes. The shifts in pulsation frequencies when these are computed from an equilibrium model where atomic diffusion is included are typically detectable with photometric data with time base of 1-year or longer. Although we have limited ourselves to stars which have a low rotation rate, atomic diffusion will most likely also be a dominant chemical element transport mechanism in faster rotating stars. A recent study by \cite{Deal2019} has found that atomic diffusion dominates over rotationally induced mixing in stars $M_\star > 1.4$\Msun for rotation velocities between 30 and 80\kms.


We have investigated the gain in constraining power when using the periods of the individual observed pulsations instead of the reduced asymptotic period spacing, \pin, as asteroseismic input \citep{Mombarg2019}. When using the individual periods, the effective temperature \teff and surface gravity \logg are used to eliminate models a posteriori that are inconsistent within the measured intervals of these observables. This study is the first to perform asteroseismic modeling of individual \gmodes in single \gDor pulsators, where we have used two slowly-rotating test stars observed by the nominal {\it Kepler} mission: \kn \citep{VanReeth2015, VanReeth2016} and \ko \citep{Kurtz2014, Takada-Hidai2017}. The use of the individual pulsations does allow us to improve upon the \xc estimates compared to the method from \cite{Mombarg2019}, but degeneracies between the mass and metallicity do not allow us to refine the mass of the star. \cart{The best model without atomic diffusion yields a mass $M_\star = 1.90$\Msun and $X_{\rm c} = 0.103$ for \kn, which are both consistent with $M_\star = 1.95 \pm 0.10$\Msun and $X_{\rm c} = 0.14^{+0.07}_{-0.09}$ from $\Pi_0$ instead of the individual mode periods.} 

We have modeled \kn and \ko from grids of stellar evolution models where the process of atomic diffusion (including radiative levitations) has been taken into account. The difference in inferred masses between models with and without atomic diffusion are typically larger than the 0.1\Msun uncertainty \cite{Mombarg2019} found from ensemble modeling of 37 \gDor stars using (\pin, \teff, \logg). Furthermore, we have investigated if the observed surface metallicities of these two stars have been altered by atomic diffusion, assuming the stars had an initial metallicity close to the solar value (i.e. fixed at $Z_{\rm ini} = 0.014$). Based on the Akaike Information Criterion (AIC), and the \chir distributions of the best 20 models, we found that models without atomic diffusion are favored in the case of \kn. For this star, we find a well-matching fit without including atomic diffusion, given the typical uncertainties on theoretically predicted periods for A/F stars. \comm{Atomic diffusion should occur in stars, hence our finding that models without it perform better but cannot fit the observed mode periods up to the measurement errors implies that other transport processes must be at work in this star.} For \ko, we found that this initial solar metallicity scenario does not yield any models consistent within the spectroscopic uncertainties derived by \cite{Takada-Hidai2017}. Yet, when the initial metallicity is set according to the spectroscopic value found by these authors, we found $M_\star = 1.36$ and $X_{\rm c} = 0.058$, which is in line with the prediction based on the average period spacing by \cite{Kurtz2014}. For this star, including atomic diffusion improves the fit to an acceptable level\comm{, without any need to consider a binary merger models -- as proposed by \cite{Takada-Hidai2017} -- from the viewpoint of g-mode asteroseismology.} 

\comm{The depth of our g-mode modeling is superior to any previous modeling of A/F-type pulsators in the literature. Moreover, we have investigated two options to choose $Y_{\rm ini}$ (when \zini is fixed at the spectroscopic value): fixing $X_{\rm ini}$ or use an enrichment rate \citep{Verma2019}. For both stars our conclusion whether including atomic diffusion improves upon the fit to the observed g-mode periods does not depend on which of these two options is chosen.}   

In this work we have investigated whether the measured surface abundances of C, O, Na, Mg, Si, S, Ca, Fe, and Ni (for \ko taken from \cite{Takada-Hidai2017}) are able to distinguish between models with and without atomic diffusion. As the predictive power of surface abundances is quite weak compared to gravity modes, we have only used these as an extra check, rather than fitting these. For \kn, this comparison was inconclusive, while for \ko the model with atomic diffusion was able to explain more of the measured abundances, although not all of them could be matched. 

The star \ko is a special case, as not all of the asteroseismic and spectroscopic findings by \cite{Kurtz2014} and \cite{Takada-Hidai2017}, respectively, can be explained. The latter study argued that \ko is most likely not a member of the thin disk, as these authors estimate the star to be located roughly 400~pc above the galactic plane, i.e. close to the edge of the thin disk. However, these estimates were based on a distance estimate from the seismic luminosity \citep{Kurtz2014}. Now that the parallax and proper motion are available from the Gaia DR2 release \citep{Gaia2016, Gaia2018}, we find that \ko is about 260~pc away from the galactic plane. Furthermore, we find a peculiar velocity of about 72\kms by using the formalism from \cite{Moffat1998}, indicating \ko might be a runaway star. Our best model with atomic diffusion is not able to explain the high observed [O/Fe] surface abundance, nor can the model with diffusion starting out at solar metallicity account for the low observed surface metallicity.

Stellar evolution models with atomic diffusion will be important for asteroseismic modeling of a large sample of \gDor stars. The tests done in this work will be repeated for the sample in \cite{VanReeth2016, VanReeth2018} in a forthcoming paper.

\acknowledgments

The research leading to these results has received funding from the European
Research Council (ERC) under the European Union’s Horizon 2020 research and
innovation programme (grant agreement N$^\circ$670519: MAMSIE) and from the
KU\,Leuven Research Council (grant C16/18/005: PARADISE), from the Research Foundation Flanders (FWO) under grant agreement G0H5416N
(ERC Runner Up Project), as well as from the BELgian federal Science Policy
Office (BELSPO) through PRODEX grant PLATO.
The computational resources and services used in this work were provided by the VSC (Flemish Supercomputer Center), funded by the Research Foundation - Flanders (FWO) and the Flemish Government – department EWI. TVR gratefully acknowledges support from the Research Foundation Flanders (FWO) through grand 12ZB620N. We thank Anne Thoul for the useful discussions on the implementation of atomic diffusion in \mesa, Sanjay Sekaran for his help with the determination of the surface abundances, and May Gade Pedersen for taking the additional spectra of \kn with the Mercator telescope. \comm{We express our gratitude towards the anonymous referee for the comments which have improved the content of this work.}
%

\vspace{5mm}
\facilities{Mercator (HERMES)}
\software{\mesa~\citep[r11701;][]{Paxton2011, Paxton2013, Paxton2015, Paxton2018, Paxton2019},~\texttt{GSSP}~\citep{Tkachenko2015},~\gyre~\citep[v5.2;][]{Townsend2013, Townsend2018}}




\newpage
\clearpage
\appendix
\section{Alternative fitting methods and initial conditions}\label{app:other_fits}
In this section we present the best models when \teff and \logg are included into the fit, which in most cases still yields models that are inconsistent within the 1-$\sigma$ intervals of these input parameters. For \ko, we also listed the best models when 1-$\sigma$ intervals for both \teff and \logg are used to eliminate models after fitting the pulsations. 
\begin{table}[htb]
    \centering
    \begin{tabular}{llll}
    \hline \hline
    \multicolumn{4}{c}{\kn}  \\
    \hline
    Spectroscopy                & Fit    & Fit    & Fit    \\ 
Diffusion                       & No     & Yes    & Yes    \\ 
$M_\star\,[{\rm M_\odot}]$      & 1.67   & 1.65   & 1.86   \\ 
$X_{\rm c}$                     & 0.293  & 0.410  & 0.242  \\ 
$X_{\rm ini}$                   & 0.715  & 0.715  & 0.715  \\ 
$Y_{\rm ini}$                   & 0.259  & 0.255  & 0.271  \\ 
$Z_{\rm ini}$                   & 0.026  & 0.030  & 0.014  \\ 
$f_{\rm ov}$                    & 0.0100 & 0.0175 & 0.0100 \\ 
$\log \chi^2_{\rm red}$         & 4.08   & 3.71   & 4.05   \\ 
$\tau\,$[Gyr]                   & 1.300  & 1.641  & 1.060  \\ 
$M_{\rm cc}\,[{\rm M_\odot}]$   & 0.177  & 0.152  & 0.194  \\ 
$\Pi_0\,$[s]                    & 4444   & 4565   & 4440   \\ 
$T_{\rm eff}\,$[K]              & 7157   & 6489   & 7563   \\ 
$\log g\,$(cgs)                 & 3.99   & 4.07   & 3.93   \\ 

    \hline

    \end{tabular}
    \caption{Best-fitting parameters for \kn when the \teff and \logg derived from high-resolution spectroscopy are included into the $\chi^2$. All parameters listed below the reduced $\chi^2$ follow from the model and are thus not free parameters. }
    \label{tab:best_models_seis+spec_975}
\end{table}

\begin{table}[htb]
    \centering
    \begin{tabular}{lllllll}
    \hline \hline
    \multicolumn{7}{c}{\ko}  \\
    \hline
Spectroscopy                    & Fit     & Fit    & Fit    & Cutoff & Cutoff  & Cutoff \\ 
Diffusion                       & No      & Yes    & Yes    & No     & Yes     & Yes    \\ 
$M_\star\,[{\rm M_\odot}]$      & 1.59    & 1.36   & 1.47   & 1.57   & 1.30    & -      \\ 
$X_{\rm c}$                     & 0.073   & 0.058  & 0.028  & 0.209  & 0.148   & -      \\ 
$X_{\rm ini}$                   & 0.715   & 0.715 & 0.715   & 0.715  & 0.715   & 0.715  \\ 
$Y_{\rm ini}$                   & 0.283   & 0.283  & 0.271  & 0.283  & 0.282   & 0.271  \\ 
$Z_{\rm ini}$                   & 0.002   & 0.002  & 0.014  & 0.002  & 0.003   & 0.014  \\ 
$f_{\rm ov}$                    & 0.0225  & 0.0225 & 0.0225 & 0.0100 & 0.0100  & -      \\ 
$\log \chi^2_{\rm red}$         & 5.09    & 5.34   & 5.47   & 6.54   & 6.52    & -      \\ 
$\tau\,$[Gyr]                   & 2.003   & 2.234  & 2.973  & 1.484  & 2.151   & -      \\ 
$M_{\rm cc}\,[{\rm M_\odot}]$   & 0.114   & 0.132  & 0.099  & 0.100  & 0.116   & -      \\ 
$\Pi_0\,$[s]                    & 3024    & 2999   & 3090   & 3086   & 2982    & -      \\ 
$T_{\rm eff}\,$[K]              & 6734    & 7622   & 6056   & 7461   & 7604    & -      \\ 
$\log g\,$(cgs)                 & 3.83    & 3.88   & 3.79   & 4.10$^\dagger$   & 4.11$^\dagger$    & -      \\ 
    \hline

    \end{tabular}
    \caption{Best-fitting parameters for \ko when the \teff and \logg for spectroscopy are included into the $\chi^2$. All parameters listed below the reduced $\chi^2$ follow from the model and are thus not free parameters. $^\dagger$Enforced to comply with the 1-$\sigma$ uncertainty interval.  }
    \label{tab:best_models_seis+spec_111}
\end{table}

\begin{figure}[htb]
    \centering
    \includegraphics[width = 0.49\textwidth]{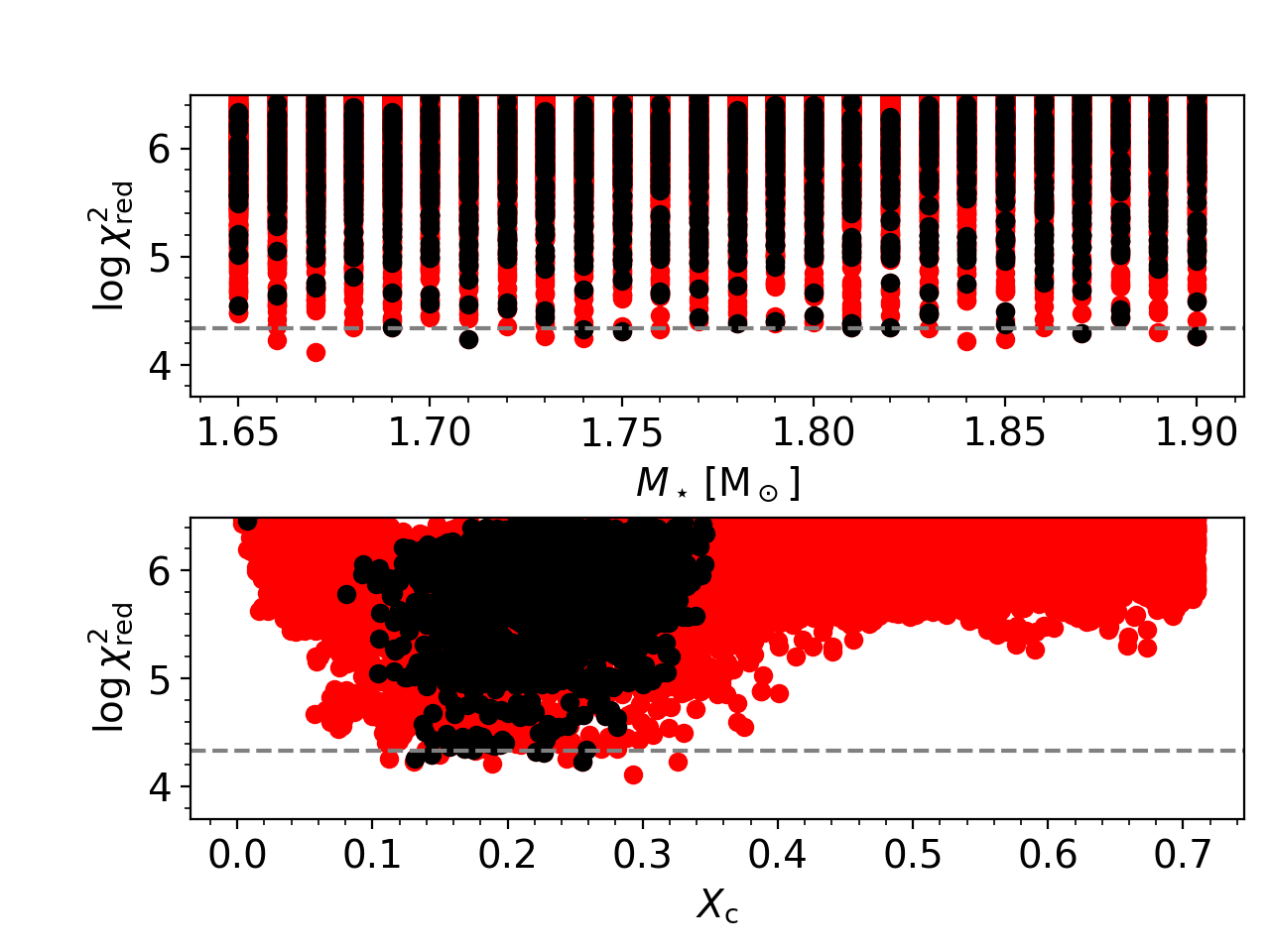}
    \caption{Reduced $\chi^2$ distributions when modeling \kn with the grid without diffusion where $Y_{\rm ini} = 1 - X_{\rm ini} - Z_{\rm ini}$. The red dots indicate the models which have been eliminated since these do not agree with the 1-$\sigma$ uncertainties on \teff and \logg. The dashed gray line indicates the 1-$\sigma$ confidence interval for the black dots. }
    \label{fig:theta-chi2_nodiff_975}
\end{figure}

\begin{figure}[htb]
    \centering
    \includegraphics[width = 0.49\textwidth]{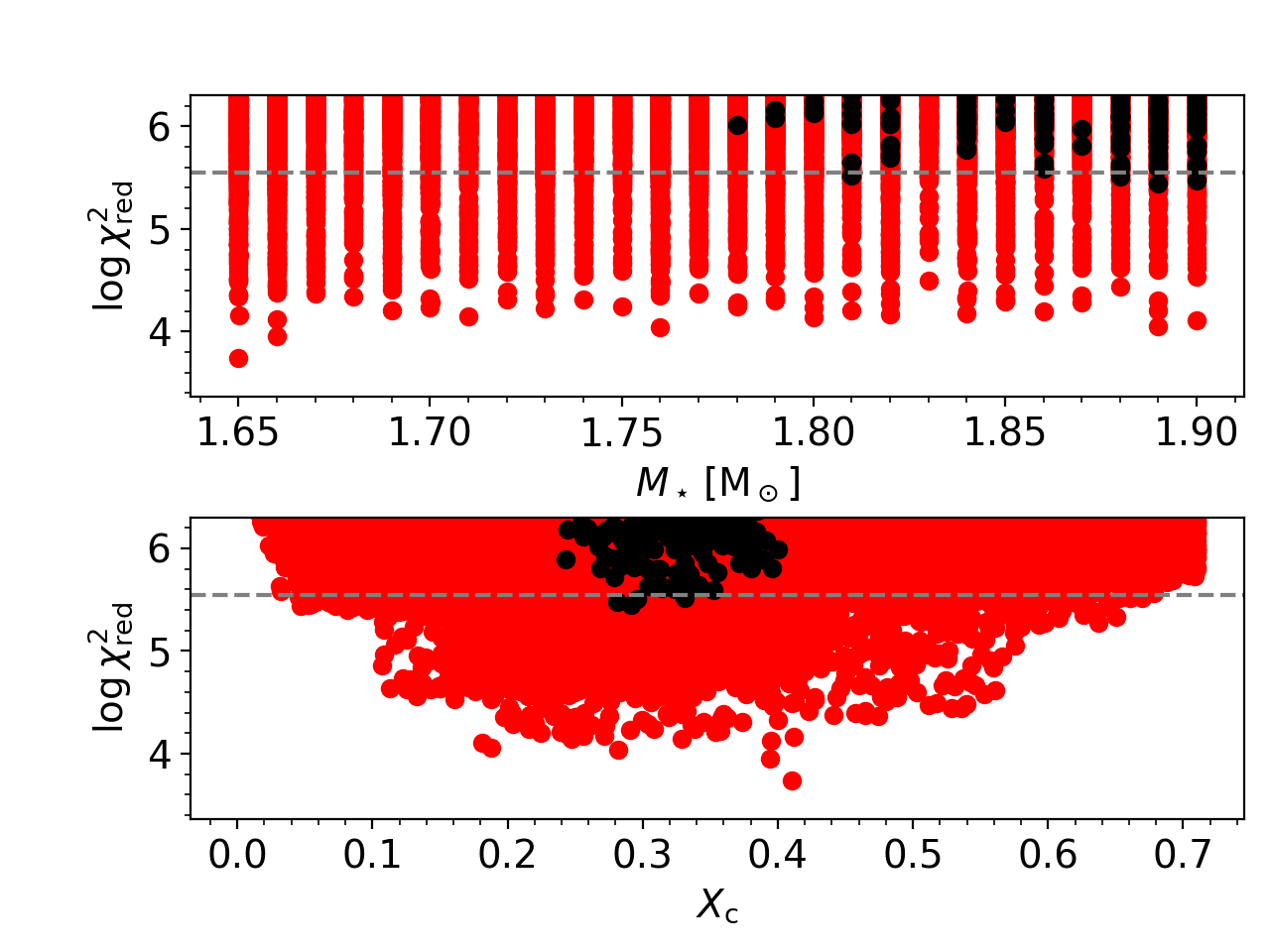}
    \caption{Same as Fig.~\ref{fig:theta-chi2_nodiff_975} (\kn), but for the grid with atomic diffusion included and a metallicity according to the spectroscopically derived value.}
    \label{fig:theta_chi2_radlev_975}
\end{figure}

\begin{figure}[htb]
    \centering
    \includegraphics[width = 0.49\textwidth]{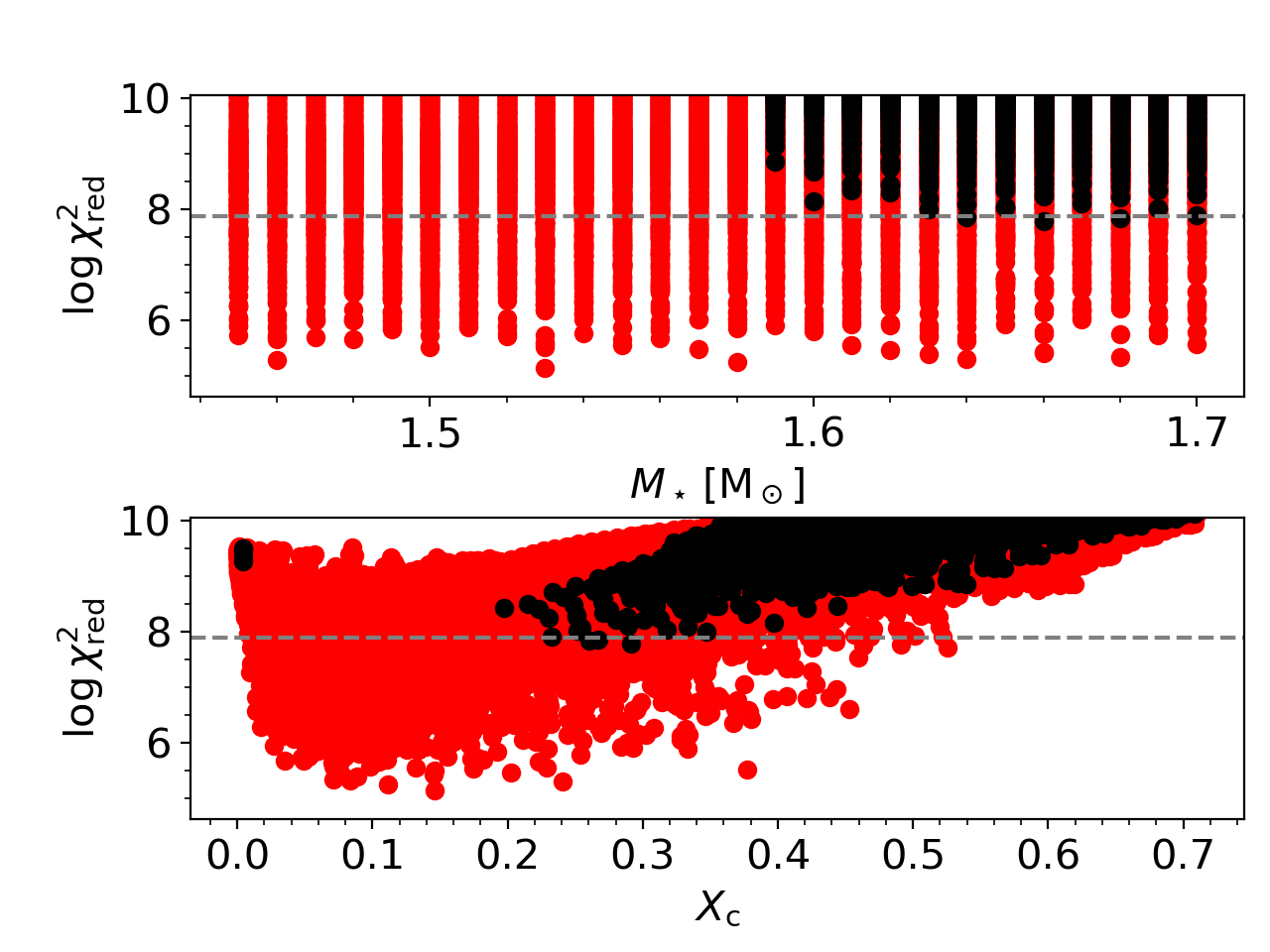}
    \caption{Reduced $\chi^2$ distributions when modeling \ko with the grid without diffusion where $Y_{\rm ini}$ is set as per \cite{Verma2019}. The red dots indicate the models which have been eliminated since these do not agree with the 1-$\sigma$ uncertainty on \teff and 3-$\sigma$ uncertainty on \logg. The dashed gray line indicates the 1-$\sigma$ confidence interval for the black dots. }
    \label{fig:theta-chi2_nodiff_111_EL}
\end{figure}

\begin{figure}[htb]
    \centering
    \includegraphics[width = 0.49\textwidth]{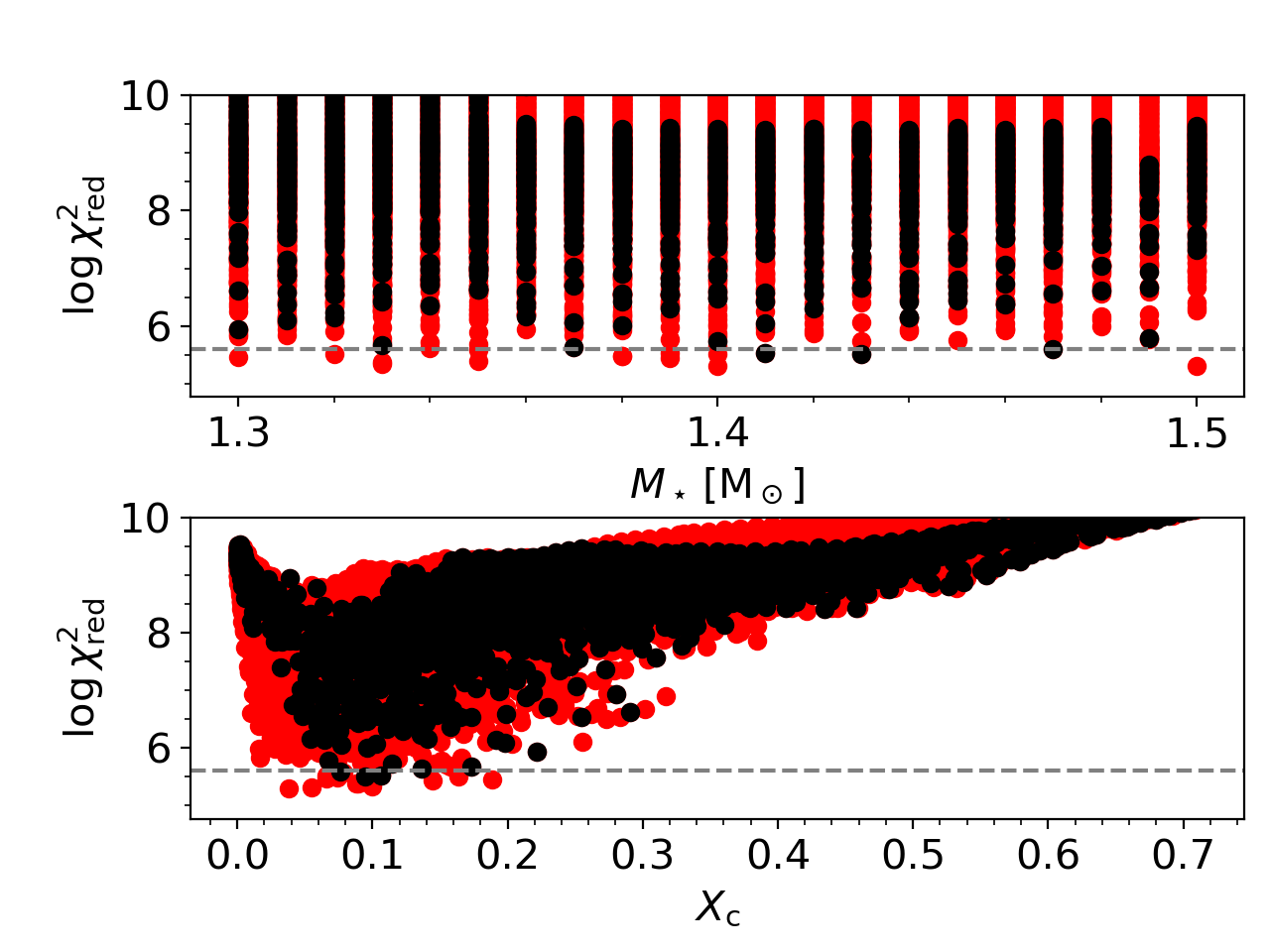}
    \caption{Same is Fig.~\ref{fig:theta-chi2_nodiff_111_EL} (\ko), but for the grid with atomic diffusion included and a metallicity according to the spectroscopically derived value.}
    \label{fig:theta_chi2_radlev_111_EL}
\end{figure}



\clearpage
\section{Conversion of abundances}\label{app:conv}
Below, we show how the abundances from \texttt{GSSP}, given in $\log(n_{\rm X}/n_{\rm tot})$, have been converted to [X/H] in order to make a comparison with the abundances predicted by the theoretical models. Firstly, the abundances are converted to $n_{\rm X}/n_{\rm H}$ using
\begin{equation}
    \log \left(\frac{n_{\rm X}}{n_{\rm tot}}\right) - \log \left(\frac{n_{\rm X}}{n_{\rm tot}}\right)_\odot = \log \left(\frac{n_{\rm X}}{n_{\rm H}}\right) + \log \left(\frac{n_{\rm H}}{n_{\rm tot}}\right) - \log \left(\frac{n_{\rm X}}{n_{\rm H}}\right)_\odot - \log \left(\frac{n_{\rm H}}{n_{\rm tot}}\right)_\odot,
\end{equation}
where $n_{\rm H}$ is the number density of hydrogen. Next, we assume $\log (n_{\rm H}/n_{\rm tot}) \approx \log (n_{\rm H}/n_{\rm tot})_\odot $, such that

\begin{equation}
    \log \left(\frac{n_{\rm X}}{n_{\rm H}}\right) = \log \left(\frac{n_{\rm X}}{n_{\rm tot}}\right) - \log \left(\frac{n_{\rm H}}{n_{\rm tot}}\right)_\odot.
\end{equation}
Finally, 
\begin{eqnarray}
    {\rm [X/H]} &=& \log \left( \frac{n_{\rm X}}{n_{\rm H}}\right) - \log \left( \frac{n_{\rm X}}{n_{\rm H}}\right)_\odot \\
    &=& \log \left( \frac{n_{\rm X}}{n_{\rm tot}}\right) - \log \left( \frac{n_{\rm H}}{n_{\rm tot}}\right)_\odot - \log \epsilon_{\rm X, \odot} + 12.
\end{eqnarray}
\bibliography{bib}

\end{document}